\documentclass{mn2e}
\usepackage{epsfig}
\usepackage{longtable}
\usepackage{mathrsfs}
\usepackage{graphicx}
\usepackage{amssymb}

\title[The protoplanetary disk Laques-Vidal 2 in Orion -- II]{Chemical abundances in the protoplanetary disk LV\,2 (Orion) -- II:
High dispersion VLT observations and microjet properties\thanks{Based on
observations made with ESO telescopes at the Paranal Observatory under
programme 078.C-0247(A).}}
\author[Y. G. Tsamis et al.]{Y. G. Tsamis$^{1, 2}$\thanks{E-mail:
ytsamis@eso.org} and J. R. Walsh$^{1}$\\
$^{1}$European Southern Observatory, Karl-Schwarzschild-Str. 2, D-85748 Garching bei M$\ddot{u}$nchen, Germany\\
$^{2}$Department of Physics and Astronomy, The Open University, Walton Hall,
Milton Keynes MK7 6AA}

\newcommand{\apj}{ApJ}
\newcommand{\apjs}{ApJS}
\newcommand{\aap}{A\&A}
\newcommand{\aj}{AJ}

\newcommand{\mnras}{MNRAS}
\newcommand{\hst}{{\it HST\/}}
\newcommand{\eld}{$N_{\rm e}$}

\newcommand{\crd}{$N_{\rm cr}$}
\newcommand{\elt}{$T_{\rm e}$}

\newcommand{\cmt}{cm$^{-3}$}

\newcommand{\cpp}{C$^{2+}$}

\newcommand{\opp}{O$^{2+}$}

\newcommand{\np}{N$^+$}
\newcommand{\sulp}{S$^+$}
\newcommand{\sulpp}{S$^{2+}$}

\newcommand{\nepp}{Ne$^{2+}$}

\newcommand{\fepp}{Fe$^{2+}$}
\newcommand{\foiii}{[O~{\sc iii}]}

\newcommand{\foi}{[O~{\sc i}]}
\newcommand{\foii}{[O~{\sc ii}]}
\newcommand{\fsii}{[S~{\sc ii}]}
\newcommand{\fsiii}{[S~{\sc iii}]}

\newcommand{\fnii}{[N~{\sc ii}]}

\newcommand{\fariv}{[Ar~{\sc iv}]}
\newcommand{\fcliii}{[Cl~{\sc iii}]}

\newcommand{\fneiii}{[Ne~{\sc iii}]}

\newcommand{\ffeiii}{[Fe~{\sc iii}]}

\newcommand{\oiii}{O~{\sc iii}}

\newcommand{\oii}{O~{\sc ii}}
\newcommand{\cii}{C~{\sc ii}}

\newcommand{\feiii}{Fe~{\sc iii}}
\newcommand{\ciii}{C~{\sc iii}}

\newcommand{\fciii}{C~{\sc iii}]}

\newcommand{\fariii}{[Ar~{\sc iii}]}

\newcommand{\hi}{H\,{\sc i}}
\newcommand{\hii}{H~{\sc ii}}
\newcommand{\hei}{He~{\sc i}}

\newcommand{\hp}{H$^+$}

\newcommand{\ha}{H$\alpha$}
\newcommand{\hb}{H$\beta$}
\newcommand{\hg}{H$\gamma$}
\newcommand{\hd}{H$\delta$}

\begin{document}

\date{Accepted ... Received ...}

\pagerange{\pageref{firstpage}--\pageref{lastpage}} \pubyear{2002}

\maketitle

\label{firstpage}

\begin{abstract}

Integral field spectroscopy of the LV\,2 proplyd is presented taken with the
VLT/FLAMES Argus array at an angular resolution of 0.31$\times$0.31 arcsec$^2$
and velocity resolutions down to 2 km\,s$^{-1}$ per pixel. Following
subtraction of the local M42 emission, the spectrum of LV\,2 is isolated from
the surrounding nebula. We measured the heliocentric velocities and widths of a
number of lines detected in the intrinsic spectrum of the proplyd, as well as
in the adjacent Orion nebula falling within a 6.6 $\times$ 4.2 arcsec$^2$ field
of view. It is found that far-UV to optical collisional lines with critical
densities, \crd, ranging from 10$^3$ to 10$^9$ \cmt\ suffer collisional
de-excitation near the rest velocity of the proplyd correlating tightly with
their critical densities. Lines of low \crd\ are suppressed the most. The
bipolar jet arising from LV\,2 is spectrally and spatially well-detected in
several emission lines. We compute the \foiii\ electron temperature profile
across LV\,2 in velocity space and measure steep temperature variations
associated with the red-shifted lobe of the jet, possibly being due to a shock
discontinuity. From the velocity-resolved analysis the ionized gas near the
rest frame of LV\,2 has \elt\ $=$ 9200 $\pm$ 800\,K and \eld\ $\sim$
10$^6$\,\cmt, while the red-shifted jet lobe has \elt\ $\approx$ 9000 --
10$^4$\,K and \eld\ $\sim$ 10$^6$ -- 10$^7$\,\cmt. The jet flow is highly
ionized but contains dense semi-neutral clumps emitting neutral oxygen lines.
The abundances of \np, \opp, \nepp, \fepp, \sulp, and \sulpp are measured for
the strong red-shifted jet lobe. Iron in the core of LV\,2 is depleted by 2.54
dex with respect to solar as a result of sedimentation on dust, whereas the
efficient destruction of dust grains in the fast microjet raises its Fe
abundance to at least 30 per cent solar. Sulphur does not show evidence of
significant depletion on dust, but its abundance both in the core and the jet
is only about half solar.


\end{abstract}


\section{Introduction}

Protoplanetary disks (proplyds) provided the first evidence of gaseous dusty
disks around young stars in the early 1990s. The archetypal proplyds found in
the Orion nebula (M42; O'Dell et al. 1993) are dense semi-ionized objects
photoevaporated to various degrees depending on their distance from the
ionizing Trapezium cluster. They are landmark objects in the study of how
circumstellar disks and eventually planetary systems form (e.g. O'Dell 2001).
They are a unique environment for the study of disk formation and evolution in
areas dominated by massive OB-type stars. Massive stellar clusters and their
associated \hii\ regions, such as in the Orion OB1 association, are thought to
represent the closest analogs to the birth environment of our Solar System
(Adams 2010). The proplyds of Orion should therefore be ideal laboratories for
the study of our planetary system's origins. At optical wavelengths proplyds
present a photoionized skin facing the ionizing cluster, giving way to a dusty
envelope often shaped into comet-shaped outflows. The elemental content and
chemistry of proplyds are \emph{very poorly} known, however, (i) the
composition of planet-forming circumstellar envelopes is of great topical
interest given the established positive correlation between host star
metallicity and the incidence of giant planetary companions (e.g. Gonzalez
1997; Neves et al. 2009); (ii) higher metallicity in the protoplanetary disk
favours the formation of giant planets in the `core accretion' scenario (e.g.
Boss 2010). While the previously discrepant abundances for the interstellar
medium and main-sequence stars in Orion are beginning to converge (Sim\'on-Diaz
\& Stasinska 2011), proplyds are the last major component of Orion for which a
chemical abundance scale is lacking.

We have been taking steps to rectify this situation with a programme targeting
a sample of bright proplyds in M42. In Tsamis et al. (2011; Paper~I hereafter)
a chemical abundance study of the protoplanetary disk LV\,2 (Laques \& Vidal
1979) and its Orion nebula host vicinity was presented, based on the analysis
of VLT optical integral field spectroscopy and \hst\ FOS single aperture
ultraviolet to far-red spectroscopy. From an emission-line analysis the
abundances of several elements were measured for the proplyd and the local M42
nebular field. LV\,2 was found to be slightly overabundant in carbon, oxygen
and neon compared to the Orion nebula gas-phase composition and to the Sun. The
carbon, oxygen and neon abundances in LV\,2 were measured to be
$\approx$0.2--0.3 dex higher than those in B-type stars of the Ori OB1
association studied by Sim\'{o}n-Diaz (2010) and Sim\'on-Diaz \& Stasinska
(2011).

That result constitutes a direct measure of the metallicity of gas
photo-evaporated from circumstellar disk material where planet formation may be
underway.

In this paper, we present higher velocity dispersion integral field spectra of
LV\,2 which enable a physical analysis of its fast microjet to be undertaken.
The paper is organized as follows. The dataset specifications and reduction
method are outlined in Section 2. The analysis and results are presented in
Section 3, followed by our conclusions in Section~4.

\section{Observations and reductions}

\setcounter{figure}{0}
\begin{figure*}
\begin{center}
\includegraphics[scale=1., bbllx=35, bblly=280, bburx=600, bbury=500, angle=0]{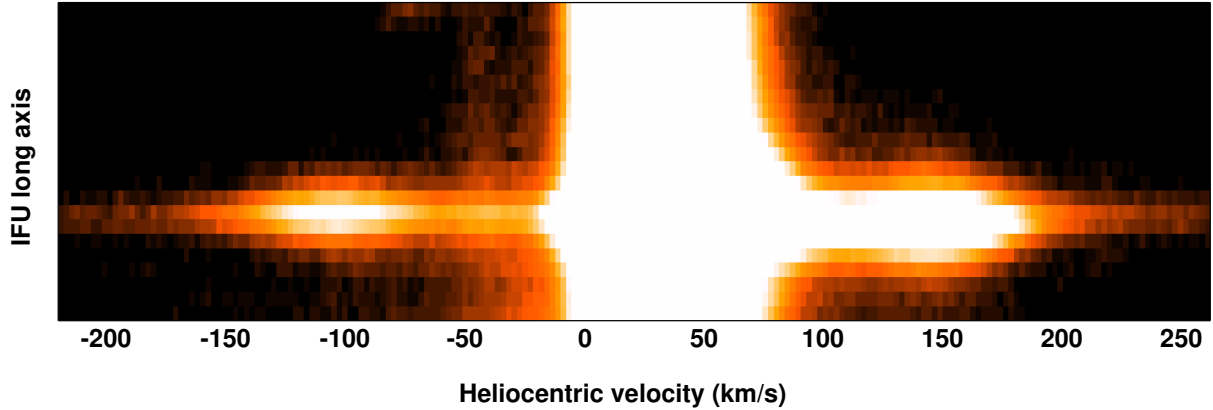}
\caption{The H$\alpha$ line observation of LV\,2 from the FLAMES HR14 grating
at a resolution of 2.29 km\,s$^{-1}$ pix$^{-1}$. The intensity scale is
logarithmic with a minimum of 5.4$\times$10$^{-17}$ and a maximum of
1.1$\times$10$^{-13}$ erg s$^{-1}$ cm$^{-2}$. The length of the vertical axis
is 6.6 arcsec (22 spaxels). The vertical spaxel size has been magnified by a
factor of 2.7 for display purposes. The Orion nebula emission has not been
subtracted. See the text for details.}

\end{center}
\end{figure*}

\setcounter{table}{0}
\begin{table}
\centering
\begin{minipage}{75mm}
\caption{Journal of VLT Argus observations.$^a$}
\begin{tabular}{@{}lcccc@{}}
\hline
              Date                   &$\lambda$-range &Grating   &$\lambda$/$\delta\lambda$   &Exp. time   \\
               (UT)                  &(\AA)           &          &   &(sec)               \\
\hline

2006/10/08                            &6383--6626      &H14B     &46\,000    &3$\times$139  \\ 
2006/12/30                            &3700--3867      &HR1      &36\,000    &3$\times$188 \\ 
2006/10/10                            &4033--4201      &HR3      &39\,000    &3$\times$185      \\ 
2006/12/30                            &4033--4201      &HR3      &39\,000    &3$\times$185    \\ 
2007/01/09                            &4188--4392      &HR4      &32\,500    &3$\times$296    \\
2007/01/09                            &4538--4759      &HR6      &32\,500    &3$\times$225 \\
2007/01/09                            &4917--5163      &HR8      &32\,000    &3$\times$130\\

\hline
\end{tabular}
\begin{description}
\item[$^a$] The Argus array was centered at (RA, Dec)$_{\rm JD2000}$ $=$
(05$^h$35$^m$16.857$^s$, $-$05$^{\circ}$23$'$15.03$''$) at a position angle of
$-$80 deg.
\end{description}
\end{minipage}
\end{table}

Integral field spectroscopy of LV\,2 was performed on the 8.2-m VLT/UT2 Kueyen
during 2006 October and December, and 2007 January with the FLAMES Giraffe
Argus array. A field of view of 6.6 $\times$ 4.2 arcsec$^{2}$ was used yielding
297 positional spectra in the optical range from six high dispersion grating
settings (Table~1). The size of the angular resolution element was 0.31
$\times$ 0.31 arcsec$^{2}$, corresponding to a spatial scale of 123 $\times$
123 AU$^{2}$ at the distance to M42 (412 pc; Reid et al 2009). The typical
seeing during the observations was $\lesssim$0.8 arcsec FWHM. The data were
cosmic ray cleaned, flat fielded, wavelength calibrated, and extracted with the
girBLDRS pipeline also employed in Paper I (see Blecha \& Simond 2004 for
details). The resulting data cubes have not been corrected for the effects of
differential atmospheric refraction (DAR; as in Paper~I), as for these high
dispersion spectra the wavelength coverage per grating is small and the effects
of DAR are negligible under the low ($\lesssim$1.5) observed air-masses.

The flux calibration was done within {\sc iraf} using exposures of various
spectrophotometric standards for the various high resolution (HR) gratings:
CD-329927 (HR1), LTT 1020 (HR3, HR8), LTT 7987 (HR14B) -- Hamuy et al. (1992),
Hamuy et al. (1994), Feige 67 (HR4) -- Oke (1990). These standards are accurate
to about one per cent for relative flux calibration and a few per cent in
absolute flux units -- under the non-photometric observing conditions, however,
the absolute flux calibration cannot be more accurate than about five per cent.
The transfer of relative flux between different wavelength ranges adds most to
the uncertainties, therefore five per cent accuracy may be a conservative
estimate.

The FWHM velocity resolution of the spectra derived from measurements of
comparison Th-Ar arc lines was 7.68 $\pm$ 0.10 km\,s$^{-1}$ at 656.3 nm (HR14B
grating), 10.1 $\pm$ 0.1 km\,s$^{-1}$ at 372.6 nm (HR1), 8.60 $\pm$ 0.20
km\,s$^{-1}$ at 409.0 nm (HR3), 9.60 $\pm$ 0.10 km\,s$^{-1}$ at 434.8 nm (HR4),
9.80 $\pm$ 0.10 km\,s$^{-1}$ at 465.8 nm (HR6), and 10.6 $\pm$ 0.1 km\,s$^{-1}$
at 496.5 nm (HR8).

\section{Results}

\subsection{Extracted spectra and LV\,2 versus M42 surface brightness}

The flux calibrated high dispersion \ha\ spectrum of LV\,2 is shown in Fig.\,~1
in the heliocentric velocity frame. The Orion nebula contribution to the line
flux has not been subtracted; the intrinsic line flux from LV\,2 is about 65
per cent of the total. This image was formed by integrating the signal of the
($X$, $Y$, $\lambda$) HR14 data cube along rows $X$= 2 -- 4 of the integral
field array which contain most of the emission from the proplyd, and then
extracting the ($-$220, $+$260) km s$^{-1}$ velocity range along columns $Y$
$=$ 1 -- 22. The proplyd peak emission is concentrated along those columns
where additional emission is detected at around $-$100 and $+$150 km\,s$^{-1}$,
respectively originating in the approaching and receding lobes of LV\,2's jet;
these are well-separated from the line systemic velocity.

\setcounter{figure}{1}
\begin{figure*}
\centering \epsfig{file=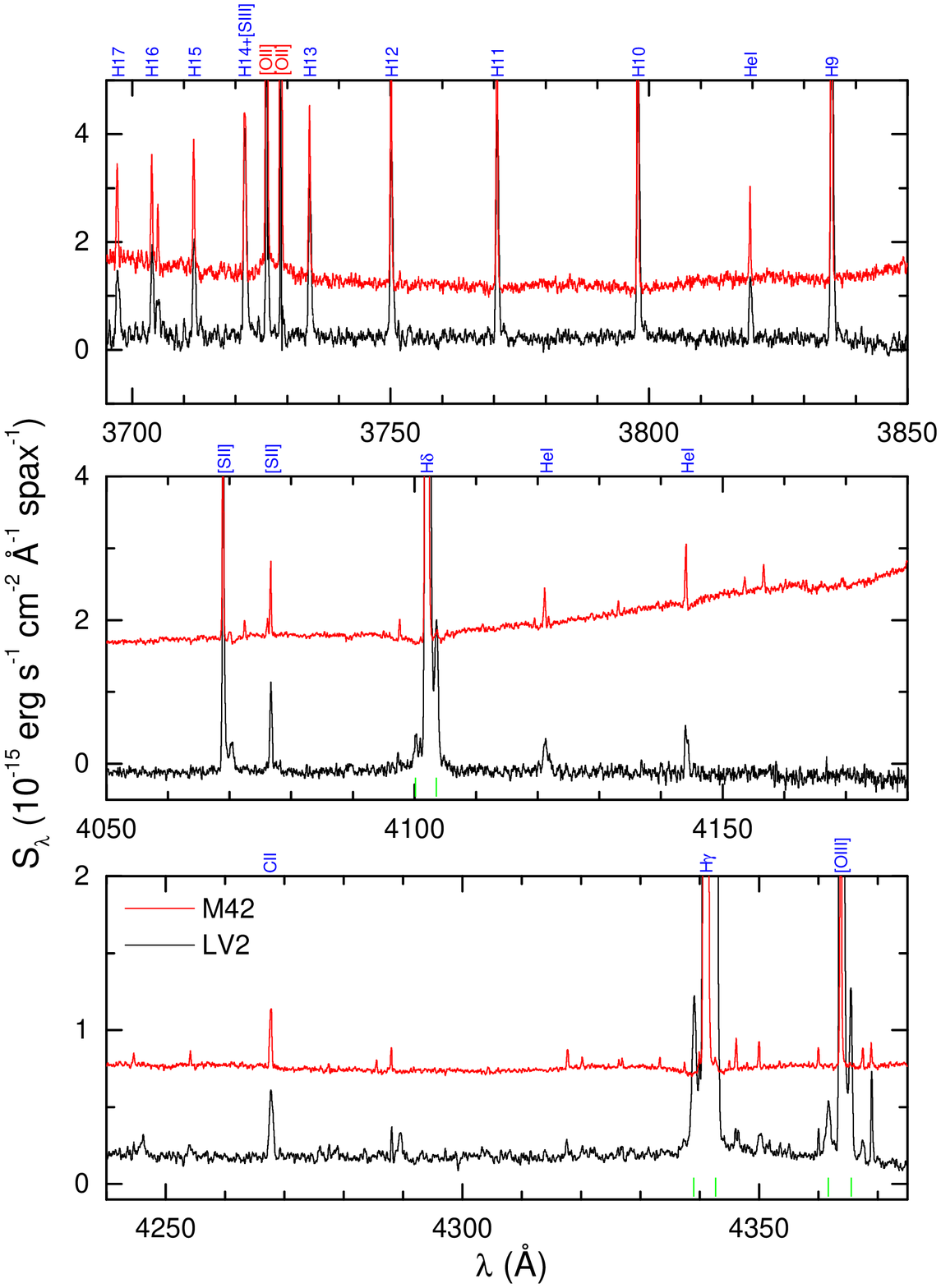, scale=.8, clip=, angle=0} \caption{Spectra of
the background subtracted LV\,2 emission (black) and local M42 vicinity (red)
from various high resolution VLT FLAMES gratings smoothed with a three point
average filter for illustration purposes (flux brightness per 0.31$\times$0.31
arcsec$^2$). Prominent lines have been identified in blue for lines common to
both LV\,2 and M42, in red for lines dominated by M42 emission. Green symbols
mark lines arising from LV\,2's bipolar jet.}
\end{figure*}

\setcounter{figure}{2}
\begin{figure*}
\centering \epsfig{file=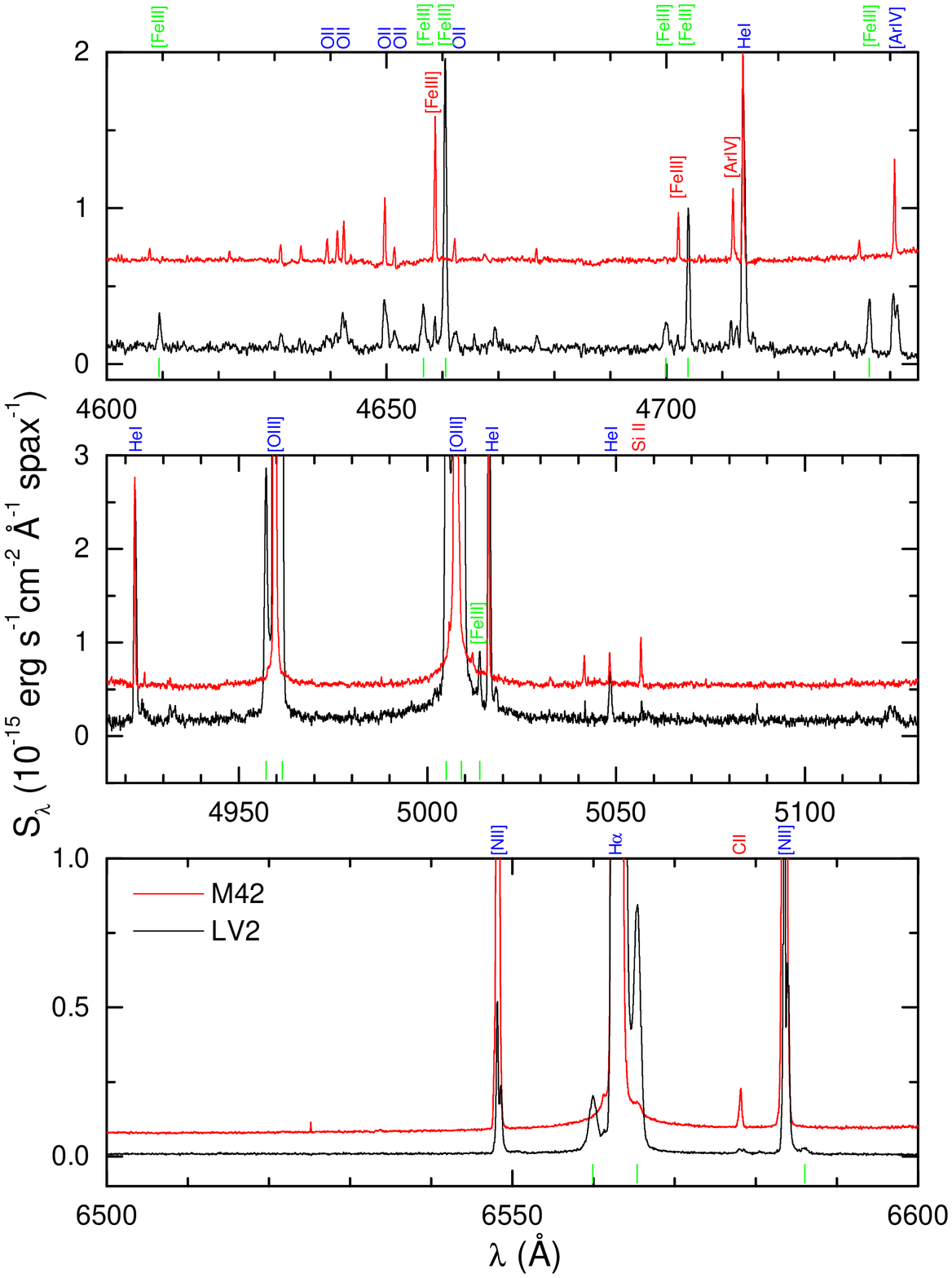, scale=.8, clip=, angle=0} \caption{As in
previous figure.}
\end{figure*}

\setcounter{figure}{3}
\begin{figure*}
\centering \epsfig{file=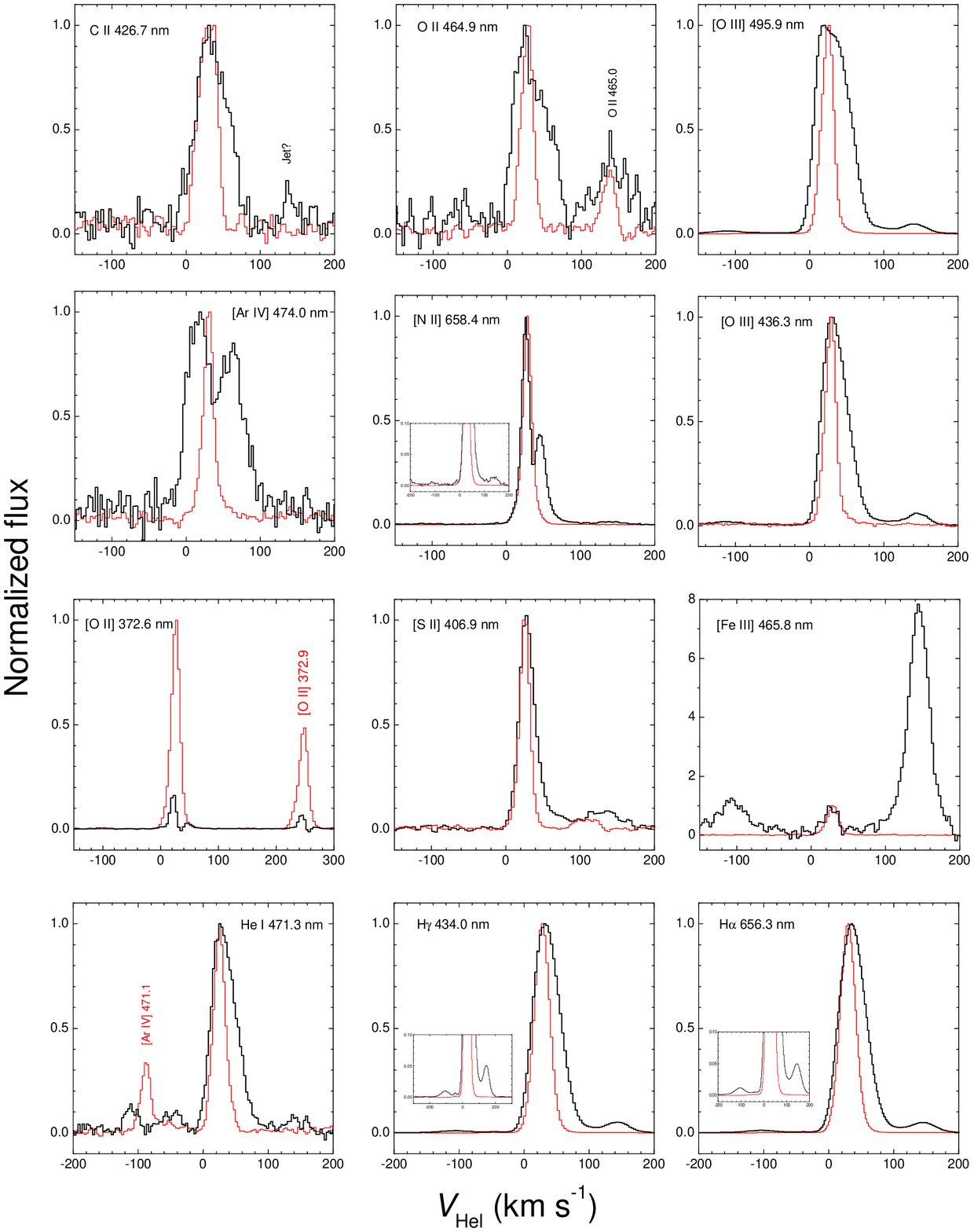, scale=.8, clip=, angle=0} \caption{A
comparison of line profiles from the intrinsic LV\,2 spectrum (black) and from
the local Orion nebula (red). The peak flux has been scaled to unity. On the
LV\,2 spectrum note the extreme weakness of \fariv\ $\lambda$4711 and of the
\foii\ $\lambda\lambda$3726, 3729 lines due to collisional de-excitation at
high densities, as well as the relative strength of the three \ffeiii\
$\lambda$4658 components.}
\end{figure*}

\setcounter{figure}{4}
\begin{figure*}
\centering \epsfig{file=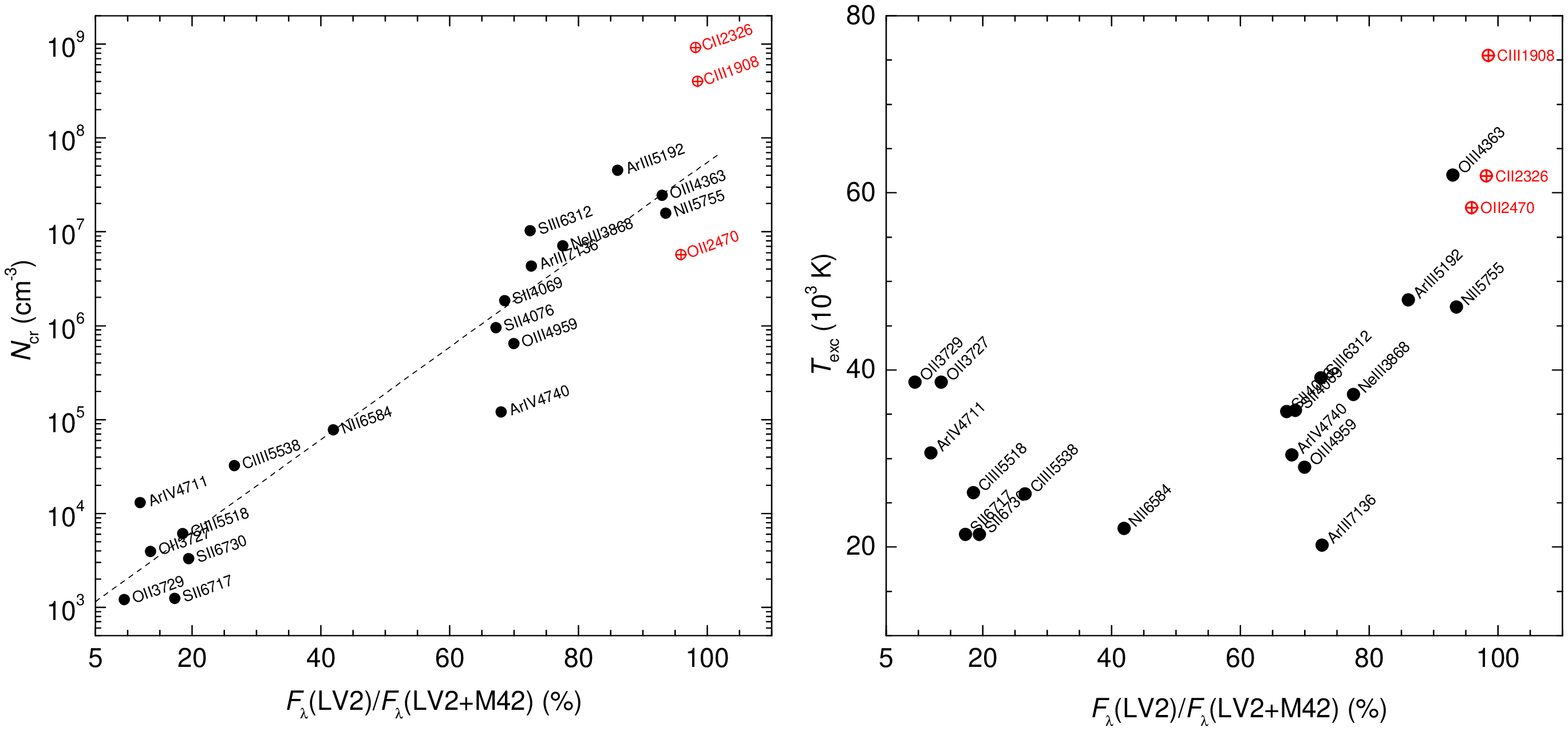, scale=.6, bbllx=0, bblly=0, bburx=450,
bbury=1400, clip=, angle=-90}
\caption{(\emph{Left}): Plot between the critical density (at 8000\,K) of
various emission lines versus the fraction (in per cent) of specific line flux
emitted from LV\,2 (background-subtracted spectrum) over that emitted from
LV\,2 $+$ M42 (observed spectrum). The red data points (crossed-circles) are
from the {\it HST} FOS measurements of Paper I and have been excluded from the
linear fit (dashed line); the rest are from the FLAMES/Argus observations.
(\emph{Right}): The excitation temperature of a given line's upper level is
plotted against the same quantity and for the same lines as in the left panel.}
\end{figure*}

Extracted 1D spectra of LV\,2 and of the local Orion nebula are shown in
Figs.\,~2 and 3. These were formed by co-adding the spectra from nine spaxels
of the reconstructed data cube containing the peak of LV\,2's emission, and
about 70 spaxels where proplyd emission was not detected (based on a \ha\ image
of the field extracted from the cube).\footnote{The typical background region
was defined as ($X$, $Y$) $=$ (4--11, 18--20) $+$ (7--12, 4--5) $+$ (11--12,
6--17) and slight variations thereof were used to test its effect on the
background-subtracted LV\,2 spectra.} The summed spectra were scaled by the
number of spaxels in each and the Orion spectrum was subtracted from the former
spectrum to obtain pure LV\,2 spectra for the various grating settings. The
local Orion nebula extraction was attempted with two different definitions of
the background, in terms of number of co-added spaxels and their location, and
the resulting LV\,2 spectral line profiles and continuum level were found to be
very similar. This simple way of performing the background subtraction and
obtaining the intrinsic proplyd emission does not take into account the fact
that LV\,2 may be more dusty than the surrounding nebula.

The issue of internal extinction in a proplyd has been discussed in detail by
Henney \& O'Dell (1999), who showed that neglecting the attenuation of flux due
to dust {\it within} a proplyd can lead to over-subtraction of the background,
in the limiting case that the principle emitting layer of the Orion nebula lies
behind a given proplyd. This can result in artificially distorted line profiles
to be obtained, for example, with double peaks and negative troughs. The
logarithmic reddening at \hb\ within LV\,2 was estimated in Paper~I to be
$\sim$0.5 dex higher than for the local Orion nebula, which translates to an
effective extinction, $A_V$, about 1 mag larger. In this data set we do not
deduce, however, a discernible wavelength-dependent effect on the resulting
line profiles that could be caused by an apparent reddening excess. For
example, the high order \hi\ lines (transitions from upper levels $n$ $>$ 10)
in the violet part of LV\,2's spectrum ($<$3800\,\AA) are all similarly shaped
and well-fitted by Gaussians (of FWHM $\sim$ 0.6\,\AA) with no evidence of
anomalous profiles. At the same time, the \foii\ violet doublet lines at 3726,
3729\,\AA\ appear heavily suppressed in the subtracted spectrum. Very weak
lines at longer wavelengths such as the recombination lines \cii\ $\lambda$4267
and \oii\ $\lambda$4649 (from recombining \cpp\ and \opp, respectively), are
well-recovered in the intrinsic proplyd spectrum (see also Fig\,~4). The \fnii\
$\lambda$6584 line is also heavily suppressed, exhibiting a central trough, a
characteristic not shared by the \fsii\ $\lambda$4069 line. In fact, the only
lines that show any evidence of altered profile shape following the background
subtraction are the forbidden lines \foii\ $\lambda\lambda$3726, 3729, \fnii\
6548, 6584, \fariv\ 4711, 4740, and perhaps only very slightly, \foiii\
$\lambda$4959 (Fig\,~4).

From this we draw the conclusion that the dominant effect on those line
profiles that appear distorted in this data set is not due to the poor
definition or over-subtraction of the Orion nebula's local emission, but that
collisionally-excited lines (CLs) are quenched near the systemic velocity of
LV\,2, due to the high plasma density there. The degree of suppression
correlates very well with the critical densities of the upper levels of the
lines in question. In Fig.\,~5 (left) the observed linear relationship is shown
between the line critical density versus the fraction, $\digamma_{\rm
\lambda}$, of specific line flux emitted from LV\,2 over the total (LV\,2 $+$
M42), normalized per spaxel. A linear fit was performed through the VLT data,
excluding for consistency the coloured data points which are from the far-UV
\hst\ FOS measurements presented in Paper~I:

\begin{equation}
{\rm log}\,N_{\rm cr} = (2.82 \pm 0.22) + (0.049 \pm 0.004)\times\digamma_{\rm
\lambda}
\end{equation}

\noindent is obtained, with a correlation coefficient of 0.96 and $\sigma$ $=$
0.46. Several data points in Fig.\,~5 (left) correspond to lines not included
in the HR grating wavelength ranges discussed here, having been obtained from
the lower resolution (LR) grating spectra of Paper~I (including \fsii\
$\lambda\lambda$6717, 6730, \fcliii, \fariii, \fneiii, and \fnii\
$\lambda$5755). In all cases the measurements are for the rest velocity
component of the lines (excluding the outlying jet lobes which correspond to
components 1 and 4 in Table 2). No correlation is evident when $\digamma_{\rm
\lambda}$ is plotted against the lines' excitation temperature (Fig.\,~5
right); this is to be expected as LV\,2 if of similar \elt\ as the local nebula
(within $\approx$800\,K). It is therefore established that due to its high
electron density ($\sim$10$^6$ \cmt), the surface brightness of the proplyd
compared to the local Orion nebula is very high in lines with large critical
densities such as the auroral \foiii\ and \fnii\ lines, and the UV transitions.
For the \oii\ $\lambda$4649 and \cii\ $\lambda$4267 lines $\digamma_{\rm
\lambda}$ $=$ 71 and 68 per cent. This is essentially identical to the value
found for \hb\ and other \hi\ Balmer lines. The significance of this result is
discussed in the conclusions.


\subsection{Line maps, velocities and widths}

The profiles of lines emitted from the proplyd as opposed to those emitted from
the local Orion nebula are compared in Fig.\,~4 in the heliocentric velocity
frame. Gaussian fits to the profiles have been employed to measure the line
heliocentric velocity and FWHM for the intrinsic proplyd spectra and for the
co-added background M42 spectrum (Table~2). Regarding the intrinsic LV\,2
spectrum, double Gaussian profiles were fitted to lines of high S/N ratio
(components 2 and 3 in Table 2 corresponding to the core or `cusp'), while
single Gaussians were fitted to the core of weaker lines. The outlying blue-
and red-shifted jet lobes were each fitted with a single Gaussian (components 1
and 4 respectively in Table 2). Lines such as \fnii\ and \fariv\ whose profiles
are double-peaked lack FWHM measurements, and their heliocentric velocities
correspond to that of their central troughs. The fit parameters pertaining to
the core of \foiii\ $\lambda$4959 are less accurate due to the increased
uncertainty of the exact line profile in the 20--40 km\,s$^{-1}$ interval
(marked by `:' in Table 2). The FWHMs in Table~2 have been corrected for
instrumental broadening by quadrature subtraction, but the thermal (Doppler)
broadening has not been corrected for: at ion temperatures of 9000\,K the
thermal width of the \hi\ and \hei\ lines is 20 and 10 km\,s$^{-1}$
respectively. The hyperfine structure causes additional broadening to lines
such as \cii\ $\lambda$4267.

The widths of lines from LV\,2 are typically larger than those from the nebula.
This is an expected result (e.g. Henney \& O'Dell 1999) and is due to the fact
that the proplyd is experiencing continuous mass-loss through a photoevaporated
flow arising from an embedded circumstellar disk by the action of far-UV
radiation (e.g. Johnstone, Hollenbach \& Bally 1998). The flow is then
converted into a photoionized wind by the Lyman continuum extreme UV (EUV)
radiation field of $\theta^1$\,Ori\,C. The ionized wind and any externally
irradiated ionized jets, when present as in LV\,2, emit essentially an \hii\
region spectrum (see Figs.\,~2, 3). The \foii, \fnii, and \fariv\ lines show
the effects of collisional quenching as discussed above. It is noteworthy that
the \cii\ and \oii\ recombination lines arising from LV\,2 appear very similar
to each other and to the \foiii\ collisional lines. In the case of \oii\ versus
\foiii\ this could mean that the lines are emitted from gas volumes of rather
similar physical properties. For example, this could be linked to the fact that
both sets of lines return similar abundances for \opp, that is, they yield a
low abundance discrepancy factor (ADF $\sim$ 1, defined as the ratio of the
recombination line (RL) over collisional line (CL)-derived \opp\ abundance; see
Paper~I). In contrast, there have been cases of planetary nebulae such as
NGC\,6153 and NGC\,7009 where the smaller widths of \oii\ RLs versus \foiii\
CLs have been linked to the presence of cool, H-poor plasma and high ADFs for
\opp\ (e.g. Barlow et al. 2006).

\setcounter{figure}{5}
\begin{figure}
\centering \epsfig{file=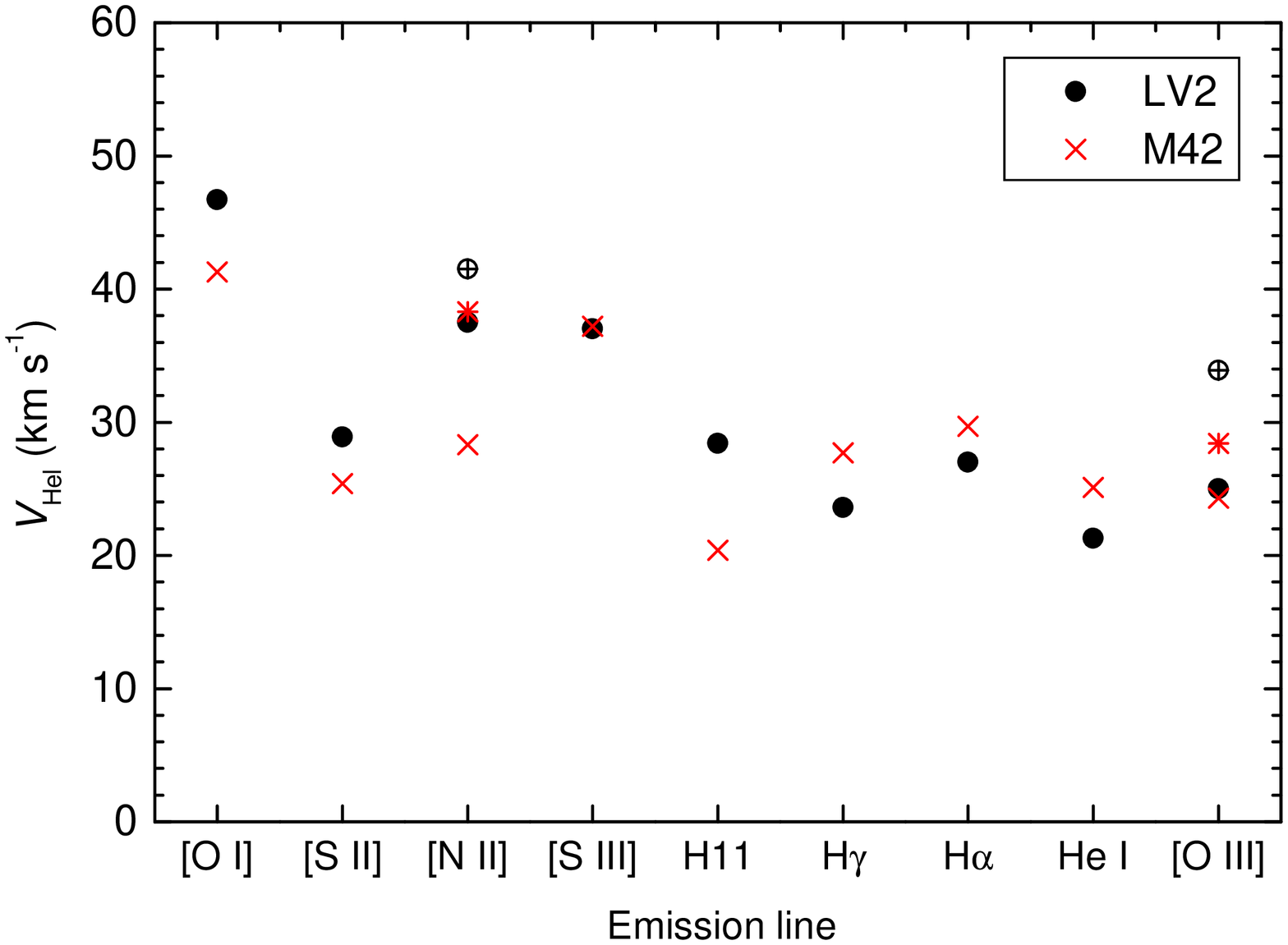, width=7.5 cm, scale=, clip=, angle=-90}
\caption{Heliocentric velocity of various lines emitted from ions of lower to
higher degree of ionization (left to right on the horizontal axis; the hydrogen
lines are all due to recombining \hp). The data for LV\,2 (black colour) were
obtained from the background-subtracted spectrum, while those for M42 (red
colour) are from the co-added spectrum as defined in the text. For \fnii\ and
\foiii\ the symbols at higher velocity correspond to the auroral lines
($\lambda$5755 and $\lambda$4363 respectively), while those at lower velocity
correspond to the nebular lines ($\lambda$6584 and $\lambda$4959). \fnii\
$\lambda$5755, \foi\ $\lambda$6300 and \fsiii\ $\lambda$6312 were measured on
the LR5 grating spectra of Paper~I.}
\end{figure}

\setcounter{figure}{6}
\begin{figure*}
\centering \epsfig{file=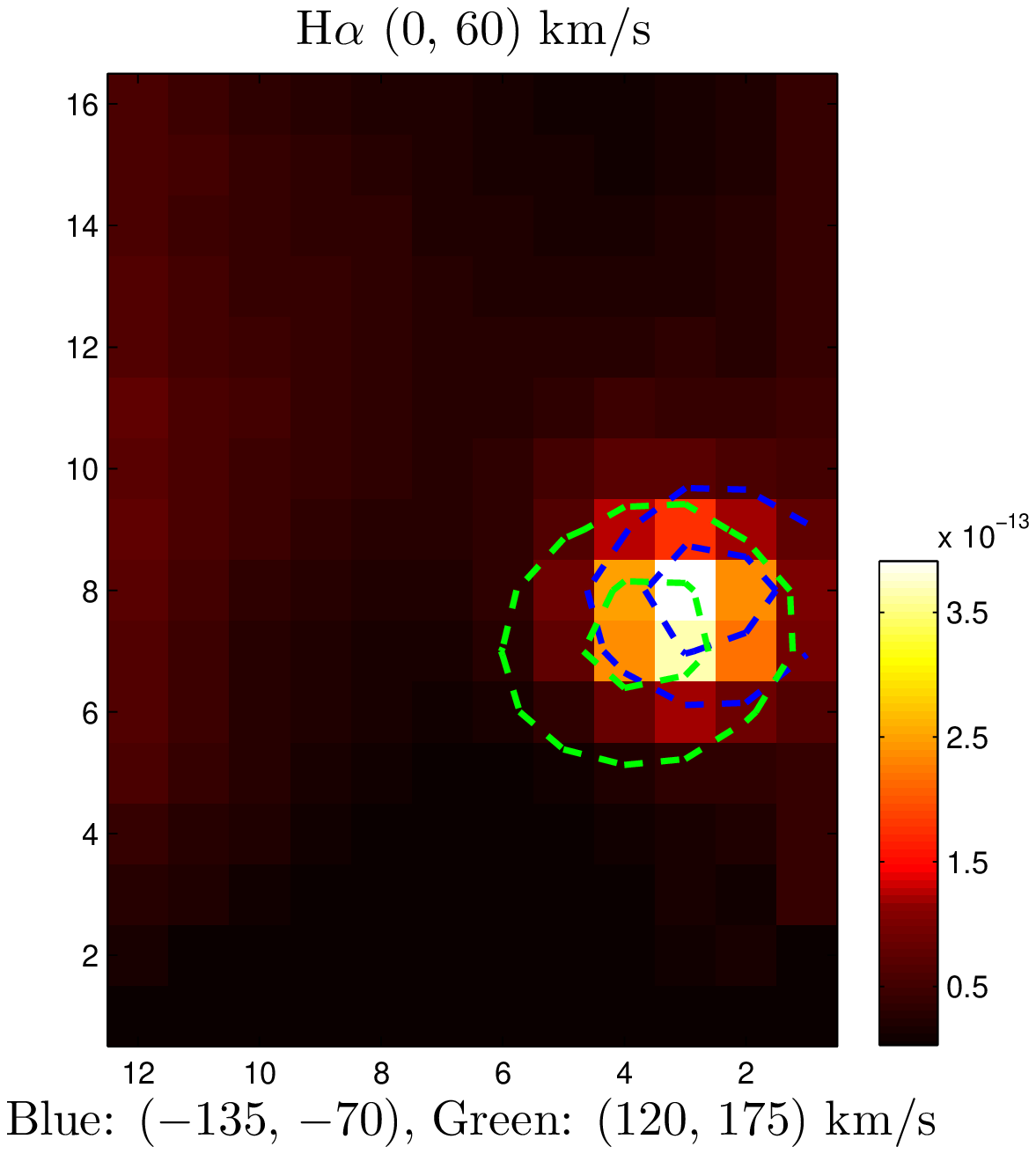, scale=0.55, clip=, bbllx=120, bblly=175,
bburx=500, bbury=675} \epsfig{file=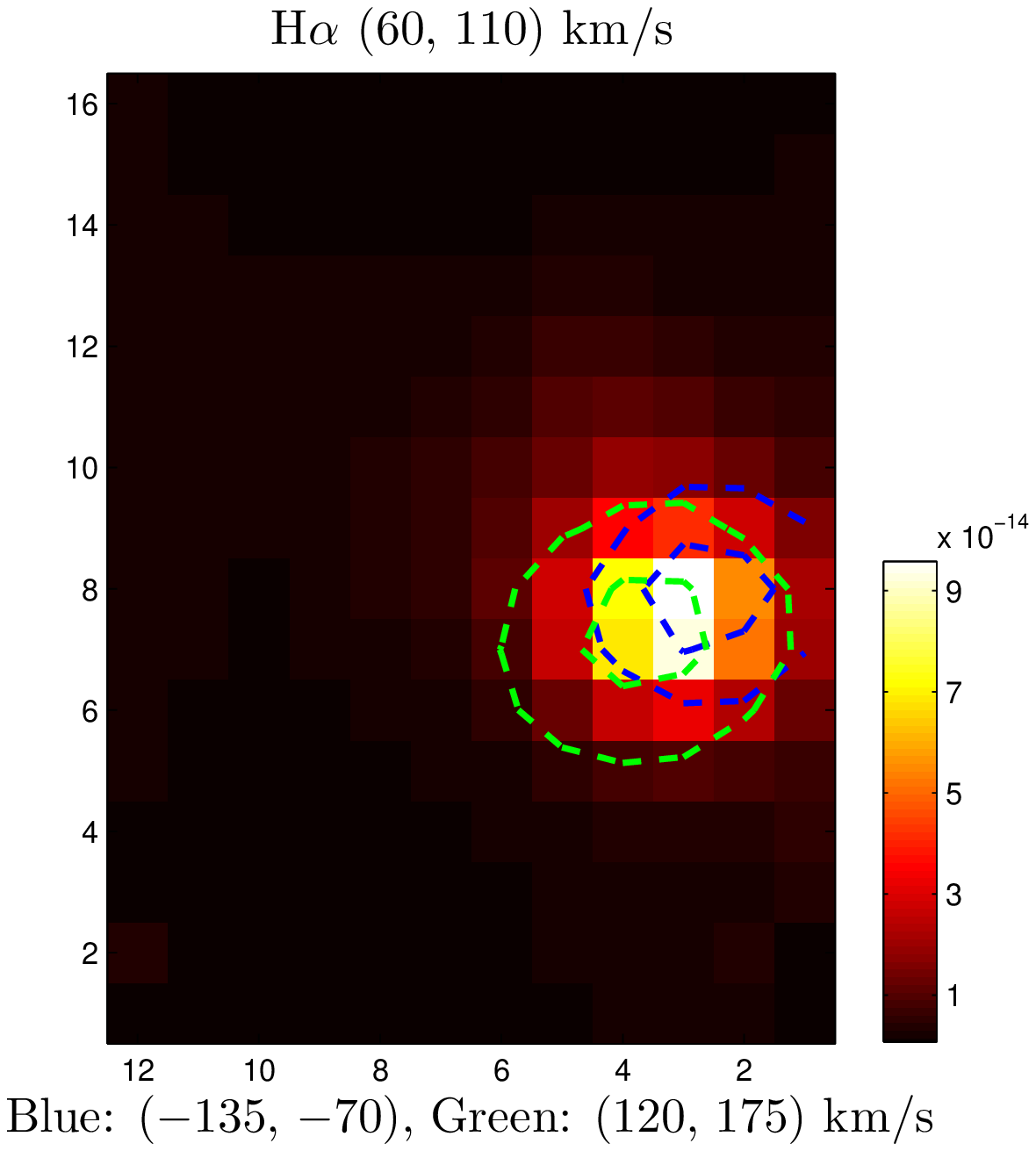, scale=0.55, clip=, bbllx=120,
bblly=175, bburx=500, bbury=675} \epsfig{file=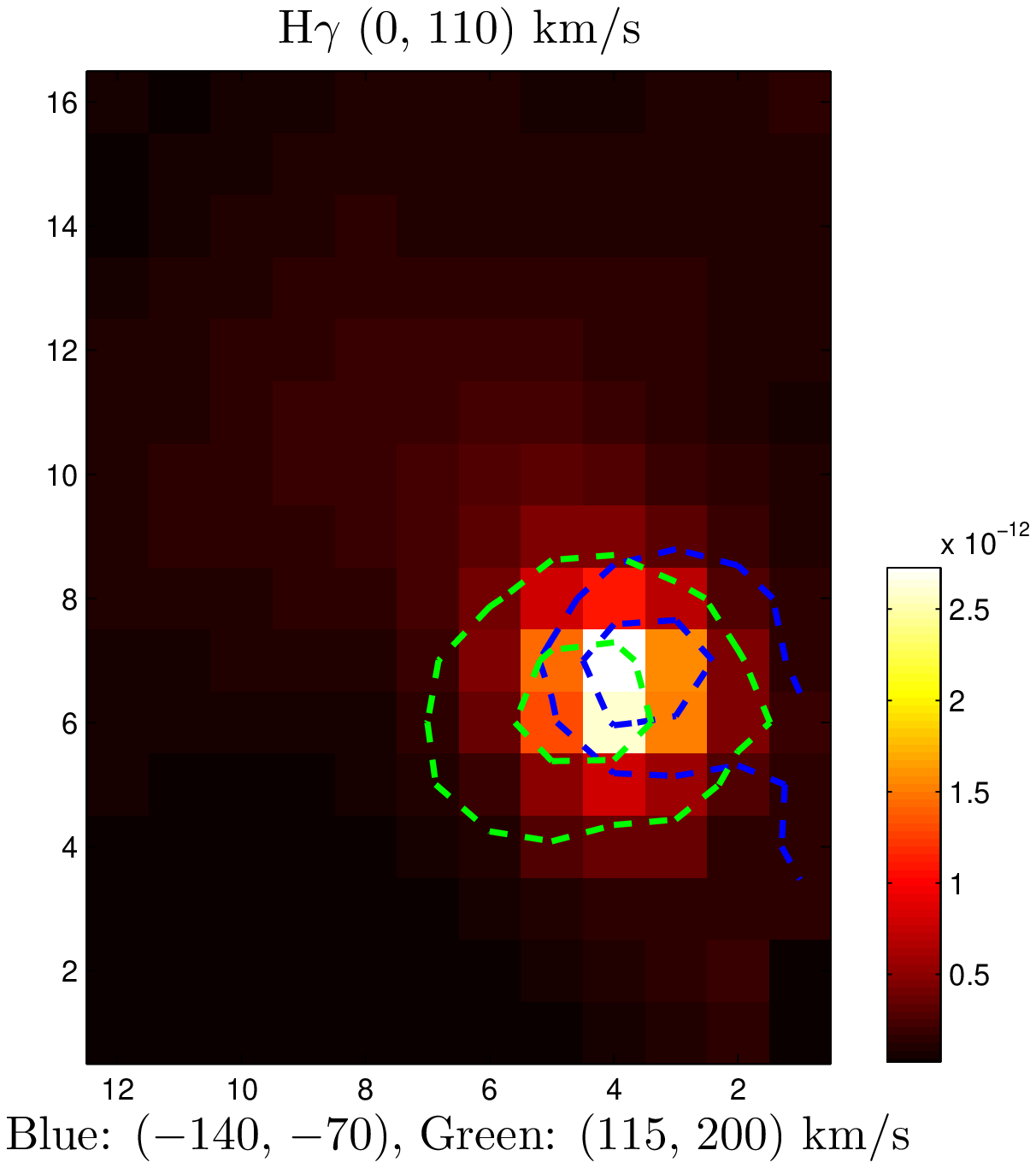, scale=0.55, clip=,
bbllx=120, bblly=175, bburx=500, bbury=675} \epsfig{file=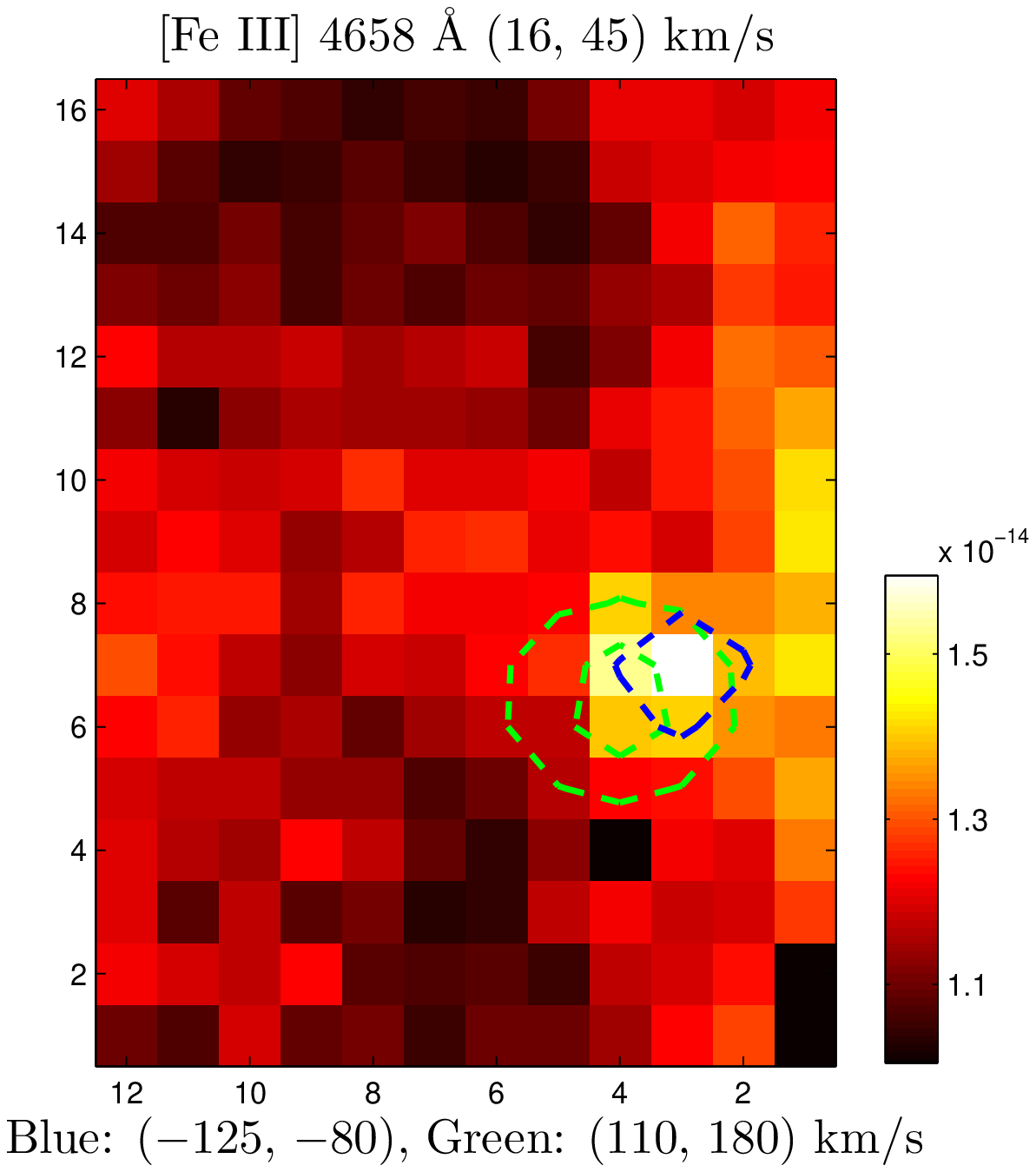, scale=0.55,
clip=, bbllx=120, bblly=175, bburx=500, bbury=675} \caption{Maps of LV\,2 and
vicinity in \hi\ and doubly ionized iron emission. Reddening correction has not
been applied. The colourmap units are erg\,s$^{-1}$ cm$^{-2}$ per 0.31$''$
$\times$ 0.31$''$ spaxel for the velocity range noted at the top of each panel.
The various contours represent emission from the proplyd's bipolar jet in
selected heliocentric velocity ranges (see bottom of each panel). The contours
in both \ha\ panels are at the 5.0$\times$10$^{-15}$, 2.0$\times$10$^{-14}$
(green) and 2.0$\times$10$^{-15}$, 5.0$\times$10$^{-15}$ (blue) levels. The
\hg\ contours are at the 2.0$\times$10$^{-14}$, 1.0$\times$10$^{-13}$ (green),
and 1.5$\times$10$^{-14}$, 3.0$\times$10$^{-14}$ (blue) levels. The \ffeiii\
contours are at the 1.0$\times$10$^{-15}$ (blue) and 1.0$\times$10$^{-14}$,
3.0$\times$10$^{-14}$ (green) levels. The \hi\ maps have been
background-subtracted. For display purposes the maps have been reduced from
their original size of 14$\times$22 spaxels.}
\end{figure*}

The velocities and FWHMs of \oii\ RLs and \foiii\ CLs in the M42 spectrum are
also very similar. The Doppler-corrected widths of \hi\ and metal lines in M42
are comparable. There is some indication for slightly larger Doppler-corrected
linewidths of \hi\ compared to \fsii\ or \foii, but not when compared to
\foiii. In an extensive study of the gas velocity structure in M42 at an
effective resolution of 10 km\,s$^{-1}$, Garc\'ia-D\'iaz et al. (2008)
concluded that the non-thermal broadening is not systematically different
between recombination lines and collisional lines. It is also notable that the
velocities of the \foii\ violet doublet components are separated by 3--4
km\,s$^{-1}$ in the M42 spectrum; the same is true for the \foiii\ nebular and
auroral transitions. This points to a degree of variation in the ion density
and/or temperature along the line of sight which warrants further
investigation. In Fig.\,~6 the behaviour of the radial velocity of lines
emitted from low to highly ionized species is examined for LV\,2 and for the
adjacent M42 nebular field. A shallow trend of decreasing velocity for
increasing ionization degree is seen for LV\,2, but the \fsii\ $\lambda$4069,
\foiii\ $\lambda$4363 and \fsiii\ $\lambda$6312 data which are due to lines of
high critical density muddle the picture. The velocities plotted are those of
component 2 (Table 2). The trend does not strengthen if instead we were to
adopt velocities obtained from single component Gaussian fits to the cusp of
LV\,2 (i.e., treating components 2 \& 3 as one component), but in that case
LV\,2 appears red-shifted versus M42 for all lines (by 6 $\pm$ 3 km\,s$^{-1}$).
The scatter of the data is larger for M42 (even amongst the \hi\ transitions
which are emitted from the \hp\ recombining ion), and this is probably real as
the observational uncertainties are typically much smaller. The behaviour of
ion velocity versus ionization potential (IP) for proplyds and their environs
has been explored by Henney \& O'Dell (1999) who found trends of increasing
blue-shift for increasing IP in three proplyds, while the opposite trend was
observed in a fourth proplyd. Regarding M42, the trend of decreasing velocity
for higher IP is commonly explained in terms of the ionization stratification
of the nebula with low ionization lines being emitted closer to the ionization
front (IF) which lies in front of the molecular cloud, and highly ionized
species existing closer to the ionizing star having been accelerated away from
the IF towards the observer (e.g. Esteban \& Peimbert 1999 and references
therein).

In Fig.\,~7 line maps of LV\,2 and its vicinity are shown, created from the
\ha, \hg, and \ffeiii\ $\lambda$4658 surface brightness distributions in
selected velocity ranges. For the \hi\ maps, sections of the field clear of
proplyd emission were fitted with Legendre polynomials and the resulting model
surfaces were subtracted from the original frames to yield
background-subtracted images. The emission from the red- and blue-shifted lobes
of LV\,2's jet is marginally spatially resolved: a separation of $\sim$1.5
spaxels (0.47 arcsec) is measured between the respective emission peaks of the
green and blue contours. The \ha\ channel maps were obtained from the highest
velocity resolution spectra of this data set and show in colour the $V_{\rm
Hel}$ $=$ 0 -- 60 and 60 -- 110 km\,s$^{-1}$ range in separate panels with the
bipolar jet emission displayed as contours. The proplyd appears more diffuse
and elongated towards the tail in the latter velocity interval which
encompasses less than 25 per cent of the line luminosity, as compared to a more
compact shape close to the line centre. This would suggest that the $\sim$60 --
110 km\,s$^{-1}$ interval intermediate between the cusp of LV\,2 and the fast
red-shifted jet lobe is occupied by gas which flows away from the cusp and goes
on to form the proplyd's comet-like tail. This corroborates the study of Henney
et al. (2002) who hinted at this possibility. The same is also evident in the
\hg\ map where the colourmap displays the whole 0 -- 110 km\,s$^{-1}$ interval.
From the \hi\ maps and a comparison between the shape of the jet contours with
the orientation of the tail it would seem that the projected axis of the
bipolar jet is at an angle of $\approx$90 deg with the tail of LV\,2.

\begin{table*}
\caption{Line structure in the intrinsic LV\,2 spectrum as compared to the
local M42 field.$^a$ }
\begin{tabular}{lcccccccc}
\noalign{\vskip3pt} \noalign{\hrule} \noalign{\vskip3pt}
&\multicolumn{8}{c}{LV\,2 proplyd}              \\

Component       &1 &2 &3 &4                 &1 &2 &3 &4     \\
                 &\multicolumn{4}{c}{$V_{\bigodot}$ (km\,s$^{-1}$)}         &\multicolumn{4}{c}{FWHM (km\,s$^{-1}$)}      \\
H12 3750    &--                   &28.7$\pm$0.4   &--         &--                     &--     &41.2$\pm$0.9  &-- &--           \\
H11 3770    &--                   &28.4$\pm$0.3   &--         &--                     &--     &44.6$\pm$0.8  &-- &--           \\
H10 3798    &--                   &28.9$\pm$0.3   &--         &--                     &--     &47.2$\pm$1.0  &-- &--           \\
H9  3835    &--                   &28.1$\pm$0.5   &--         &--                     &--     &46.1$\pm$0.7  &-- &--           \\
\fsii\ 4069    &--                &28.8$\pm$0.3   &--         &129.8$\pm$1.4          &--     &24.0$\pm$0.8  &-- &52.7$\pm$4.8 \\
\fsii\ 4076    &--                &28.7$\pm$0.5   &--            &--          &--     &36.2$\pm$0.8  &-- &-- \\
H$\delta$      &--                &19.0$\pm$0.5   &46.1$\pm$0.6   &133.9$\pm$0.5        &-- &34.4$\pm$0.8 &34.8$\pm$0.8 &48.7$\pm$1.2      \\
\cii\ 4267    &--                 &34.8$\pm$0.9   &--                   &--              &--                 &52.6$\pm$2.0   &--             &--           \\
H$\gamma$      &$-$105.6$\pm$0.6  &23.6$\pm$0.3   &41.6$\pm$1.3   &141.8$\pm$0.5    &62.1$\pm$1.7       &30.9$\pm$1.2   &45.1$\pm$0.9   &47.4$\pm$1.0           \\
\foiii\ 4363   &$-$112.7$\pm$5.5  &24.7$\pm$0.5   &41.9$\pm$2.8   &143.8$\pm$0.5   &73.9$\pm$14.4      &22.3$\pm$1.8   &36.3$\pm$2.6   &38.6$\pm$1.1            \\
\ffeiii\ 4607  &--                &--             &--     &150.8$\pm$1.0   &--                 &--             &--     &37.1$\pm$2.5                     \\
\oii\ 4649    &--                 &20.6$\pm$0.6   &54.2$\pm$1.1   &--       &--         &28.0$\pm$2.7   &34.5$\pm$5.4   &--                     \\
\oii\ 4650    &--                 &32.3$\pm$3.2   &--             &--     &--                 &69.0$\pm$8.0   &--             &--                       \\
\ffeiii\ 4658  &$-$106.9$\pm$2.1  &26.4$\pm$1.7   &--     &144.9$\pm$0.2     &58.6$\pm$5.0       &25.6$\pm$3.9   &--     &33.1$\pm$0.5                \\
\oii\ 4661    &--                 &32.8$\pm$2.4   &--             &--      &--                 &62.2$\pm$6.0   &--             &--                     \\
\ffeiii\ 4702  &$-$114.6$\pm$2.0  &17.5$\pm$2.0   &--     &139.4$\pm$0.2    &69.0$\pm$5.3       &25.7$\pm$4.7   &--     &32.0$\pm$0.6                   \\
\hei\ 4713    &--         &21.3$\pm$0.6   &39.7$\pm$3.9   &--        &--         &24.9$\pm$2.9   &41.6$\pm$3.5   &--                           \\
\fariv 4740   &--         &35.2:     &--     &--            &--         &--     &--     &--                                                  \\
\hei\ 4922    &--         &30.3$\pm$0.5   &--             &--         &--         &46.9$\pm$1.3   &--             &--                            \\
\foiii\ 4959   &$-$109.0$\pm$2.0     &25:     &55:        &138.8$\pm$0.5     &45.2$\pm$2.3       &36:             &28:        &42.1$\pm$1.1                         \\
\ffeiii\ 5011  &--                 &--             &--     &142.1$\pm$0.5        &--                 &--             &--     &35.8$\pm$1.1              \\
\fnii\ 6584    &--         &39:   &--            &139.0$\pm$1.0   &--         &--         &--         &39.6$\pm$2.0       \\
H$\alpha$  &$-$102.9$\pm$2.2             &27.0$\pm$2.1           &43.2$\pm$7.4     &142.7$\pm$0.5        &51.9$\pm$1.5               &36.7$\pm$6.7           &47.5$\pm$3.6     &50.0$\pm$3.0              \\

&\multicolumn{8}{c}{Local M42 nebula}              \\
        &\multicolumn{4}{c}{$V_{\bigodot}$ (km\,s$^{-1}$)}         &\multicolumn{4}{c}{FWHM (km\,s$^{-1}$)}       \\
\foii\ 3727    &--              &25.0$\pm$0.2   &--         &--                     &--     &14.3$\pm$0.3          &-- & --          \\
\foii\ 3729    &--              &21.8$\pm$0.3   &--         &--                     &--     &15.7$\pm$0.4          &-- & --          \\
H12 3750    &--                 &21.0$\pm$0.3   &--         &--                     &--     &25.8$\pm$0.3          &-- & --          \\
H11 3770    &--                 &21.0$\pm$0.2   &--         &--                     &--     &26.0$\pm$0.5          &-- & --          \\
H10 3798    &--                 &20.3$\pm$0.4   &--         &--                     &--     &26.5$\pm$0.5          &-- & --          \\
H9  3835    &--                 &20.3$\pm$0.3   &--         &--                     &--     &25.4$\pm$0.4          &-- & --          \\
\fsii\ 4069    &--              &25.4$\pm$0.3   &--    &--          &--     &12.5$\pm$0.6  &-- &-- \\
\fsii\ 4076    &--              &25.4$\pm$0.5   &--       &--          &--     &11.5$\pm$0.8  &-- &-- \\
H$\delta$         &--           &21.8$\pm$0.1   &--     &--             &--     &25.4$\pm$0.4  &-- &-- \\
\cii\ 4267        &--           &29.3$\pm$0.5   &--             &--          &--                 &27.8$\pm$1.3   &--             &--             \\
H$\gamma$         &--           &27.7$\pm$0.0   &--     &--                 &--         &26.0$\pm$0.1   &--     &--                           \\
\foiii\ 4363        &--         &28.4$\pm$0.2   &--     &--             &--         &15.3$\pm$0.5   &--     &--                               \\
\ffeiii\ 4607  &--              &34.6$\pm$0.6   &--     &--        &--                 &16.9$\pm$1.4   &--     &--                           \\
\foii\ 4649        &--          &26.8$\pm$0.2   &--     &--     &--         &16.6$\pm$0.5   &--     &--                                     \\
\foii\ 4650        &--          &26.6$\pm$0.4   &--     &--     &--         &18.6$\pm$1.0   &--     &--                                  \\
\ffeiii\ 4658        &--        &29.1$\pm$0.2   &--     &--     &--         &15.9$\pm$0.5   &--     &--                                \\
\oii\ 4661        &--           &24.9$\pm$0.6   &--     &--     &--         &18.6$\pm$1.4   &--     &--                                \\
\ffeiii\ 4702  &--              &25.5$\pm$0.2   &--     &--         &--                 &14.6$\pm$0.6   &--     &--                    \\
\hei\ 4713        &--           &25.1$\pm$0.2   &--     &--     &--         &20.1$\pm$0.3   &--     &--                                \\
\fariv\ 4740        &--         &30.4$\pm$0.3   &--     &--     &--         &14.9$\pm$0.7   &--     &--                                  \\
\hei\ 4922        &--           &22.7$\pm$0.1   &--     &--     &--         &18.9$\pm$0.3   &--     &--                                \\
\foiii\ 4959        &--         &24.3$\pm$0.2   &--     &--     &--         &16.2$\pm$0.4   &--     &--                                  \\
\ffeiii\ 5011  &--              &26.1$\pm$0.4   &--     &--         &--                 &12.0$\pm$0.9   &--     &--                    \\
\fnii\ 6584    &--              &28.3$\pm$0.2   &--            &--      &--         &11.1$\pm$0.4       &--         &--         \\
H$\alpha$ &--                   &29.7$\pm$0.1           &--     &--                   &--                 &26.5$\pm$0.3           &--     &--                        \\

\noalign{\vskip3pt} \noalign{\hrule}\noalign{\vskip3pt}
\end{tabular}
\begin{description}
\item[$^a$] Velocities are heliocentric. The FWHMs have been corrected for
instrumental broadening using the Th-Ar arc line measurements reported in
Section 2. Highly uncertain values are marked with `:'.
\end{description}
\end{table*}

\subsection{Physical conditions and jet properties}

The red-shifted lobe of LV\,2's jet is prominent in several emission lines from
hydrogen, helium and heavier species (O$^0$, \np, \sulp, \sulpp, \opp, \nepp,
\fepp). The blue-shifted lobe is also well-detected in [O~{\sc i}], \foiii,
\fsiii, \ffeiii $\lambda$4658, and the strong \hi\ lines (see Figs.\,~2, 3 of
this paper and fig.\,~7 of Paper~I): the lines are however a factor of a few
weaker. A compilation of line fluxes from the high-dispersion HR spectra of
this paper and the lower-dispersion LR spectra of Paper~I is presented in
Table~3 for the red-shifted lobe.

In order to compensate for the fact that there are gaps in the wavelength
coverage of the FLAMES HR gratings and that the LR and HR observations were
taken on different nights and under different observing conditions, the HR
measurements have been cast in units relative to \hb\ using the LR spectra of
Paper~I in the following manner:

\begin{equation}
\frac{F(\lambda)_{\rm jet}}{F(\rm H\beta)_{\rm jet}} = \big
[\frac{F(\lambda)_{\rm jet}} {F(\lambda)_{\rm core}}\big]_{\rm HR} \times
\big[\frac{F(\lambda)_{\rm core}}{F(\rm H\beta)_{\rm jet}}\big]_{\rm LR}.
\end{equation}

\noindent The \ffeiii\ $\lambda$4658 line was scaled via \hei\ $\lambda$4713,
and \ffeiii\ $\lambda$5011 was scaled via \hei\ $\lambda$5016. \ha\ was scaled
via \fnii\ $\lambda$6584 as the rest velocity component of \ha\ probably
suffers from saturation effects on the HR14 and LR5/6 grating spectra. The
relative line fluxes were then dereddened with $c$(\hb) $=$ 1.2$\pm$0.2 using
the modified Cardelli, Clayton \& Mathis (1989; CCM) law from Blagrave et al.
(2007), with a total to selective extinction ratio $R_{\rm V}$ $=$ 5.5. This
reddening constant was obtained from the H$\delta$/\hb\ and H$\gamma$/\hb\
ratios from a comparison of the observed ratios to their theoretical values
from Storey \& Hummer (1995) at \elt\ $=$ 10$^4$\,K and \eld\ $=$ 10$^6$ \cmt.
The observed \ha/\hb\ ratio is considered unreliable due to \ha\ likely being
saturated. Using the same \hi\ line ratios, $c$(\hb) for the blue-shifted jet
lobe is measured to be 0.9 and 1.5 respectively, in approximate agreement with
the reddening deduced for the red-shifted lobe above and for the cusp of LV\,2
from Paper~I.

In Fig.\,~8, a group of curves is shown on the (\elt, \eld) diagnostic plane
representing solutions for the dereddened line ratios from Table~3 applicable
to the red-shifted jet. The ratios \fnii\ $\lambda$5755/$\lambda$6584, \foi\
$\lambda$5577/$\lambda$6300, \foiii\ $\lambda$4363/$\lambda$5007, and \ffeiii\
$\lambda$5011/$\lambda$4658 ($^3$P$_1$--$^5$D$_2$)/($^3$F$_4$--$^5$D$_4$) were
used, employing the multi-level ion modelling program {\sc equib} developed at
University College London. The \ciii] $\lambda$1907/$\lambda$1909 curve is from
the {\it HST} STIS observations of LV\,2 by Henney et al. (2002). From Fig\,~8
we deduce that the mean electron temperature and density of the jet are 9000 --
10,000\,K and $\sim$10$^6$ \cmt, respectively. The iron lines indicate that
overall higher densities of 2.5$\times$10$^6$ \cmt\ are appropriate for their
emitting region, and densities log (\eld/\cmt) $=$ 7.0$^{+1.0}_{-0.6}$ are
obtained from the \ffeiii\ $\lambda$4702/$\lambda$4658
($^3$F$_3$--$^5$D$_3$)/($^3$F$_4$--$^5$D$_4$) ratio when comparing with the
theoretical ratios of Keenan et al. (2001). The ionized carbon, nitrogen, and
neutral oxygen lines also indicate densities of $\sim$10$^6$ \cmt. The fact
that a range of densities are obtained from various diagnostics is an
indication that the jet is stratified in density and probably has a clumpy
small-scale structure. This is corroborated by the presence of neutral oxygen
in the body of the jet, already noted in Paper I (see fig.\,~7), indicating
that dense quasi-neutral clumps can withstand complete photoionization in the
jet flow.

Our analysis suggests that previous mass-loss rate estimates from the jet of
$\sim$10$^{-8}$ M$\odot$ yr$^{-1}$, which were based on considerations of a
fully ionized flow (e.g. Henney et al. 2002; Vasconcelos et al. 2005), would
actually be lower limits.

\subsubsection{Velocity-resolved physical conditions}

We used the high dispersion \foiii\ line profiles obtained with the VLT Argus
fibre array in conjunction with the \ciii] {\it HST} STIS observations by
Henney et al. (2002) from programme 8120 to map the variation of electron
temperature and density as a function of the heliocentric velocity of the
ionized gas in LV\,2. These \fciii\ and \foiii\ lines should originate from
roughly the same volume of highly ionized gas. The results from this analysis
are shown in Fig.\,~9 where the \foiii\ profiles are from the intrinsic
(background-subtracted) LV\,2 spectrum. The \foiii\ and \ciii] profiles were
first binned to a velocity resolution of 3 km\,s$^{-1}$ pix$^{-1}$ and the
electron density was computed as a function of $V_{\rm hel}$ from the
density-sensitive (reddening and \elt-independent) \ciii]
$\lambda$1907/$\lambda$1909 ratio for an electron temperature of 9000\,K. The
obtained density profile (Fig.\,~9 middle panel) was then used to compute the
temperature profile from the \foiii\ auroral to nebular line ratio (Fig.\,~9
top panel). The density variations that are observed across the bulk of LV\,2's
emission in the 10--80 km\,s$^{-1}$ range are essentially the same as those
obtained by Henney et al. (2002) who analyzed the \ciii] data in a similar way.
The electron density in LV\,2 peaks very close to the rest velocity of the
\foiii\ lines [log\,\eld\ (cm$^3$) $=$ 6], then falls off towards a local
minimum at $\sim$100 km\,s$^{-1}$, before rising again to values of log\,[\eld\
(cm$^3$)] $=$ 6 and higher in the velocity range occupied by the red-shifted
jet emission.

The analysis reveals that steep temperature variations are observed: (i) in the
velocity interval between the bulk LV\,2 emission and the onset of the
red-shifted jet lobe, (ii) at higher velocities beyond the `red' edge of the
jet. The observed amplitude is $\simeq$3500\,K with a maximum of 12\,500\,K on
the `blue' side of the jet, while even higher temperatures (admittedly with
very large associated uncertainties) are observed on its `red' side.  These
steep \elt\ variations could be the signature of a shock discontinuity and its
associated heating/cooling zone which is kinematically resolved here. A shock
discontinuity can form at the locus where the jet encounters and bursts out
from the main ionization front surrounding LV\,2 (the H$^0$/\hp\ contact
surface) and/or at the outer physical envelope of the conical/cylindrical jet.

On the other hand, at velocities associated with the bulk emission near the
rest frame of LV\,2, and with the bulk jet emission, the temperature is
$\sim$9000\,K which is typical of photoionized gas. This measurement (the
horizontal black line in Fig.\,~9 top) is corroborated by the analysis of the
integrated line ratios for the rest velocity component (9000 $\pm$ 600\,K) from
the lower-dispersion spectra of Paper~I (table 3), and from the diagnostic
diagram pertaining to the jet in the present work (Fig.\,~8). In the
temperature plot of Fig.\,~9 the red data points correspond to the intrinsic
\foiii\ line profiles where the interval 20--40 km\,s$^{-1}$ has been excluded
from the analysis due to the increased uncertainty in the exact profile of
$\lambda$4959 very close to the bulk velocity of M42. Specifically, \elt\ $=$
9240 with $\sigma$ $=$ 815\,K for $V_{\rm Hel}$ $\in$ [0, 80] km\,s$^{-1}$
across LV\,2's cusp (where the scatter indicates real variations rather than
observational uncertainties). This velocity interval contains $\gtrsim$98 per
cent of the intensity of the \foiii\ auroral and nebular lines near the rest
frame of LV\,2, excluding the outlying jet lobes. The analysis was also
performed using the observed (LV\,2 $+$ M42) line profiles resulting in a
variation (black crosses) that follows closely the previous one across most of
the velocity range sampled. In this case we find that across the cusp \elt\ $=$
9020 with $\sigma$ $=$ 910\,K for $V_{\rm Hel}$ $\in$ [0, 80] km\,s$^{-1}$. We
therefore deduce that the bulk of the \foiii\ emitting region in LV\,2 appears
to be fairly isothermal with a $\lesssim$10 per cent rms temperature variation,
while considering the uncertainties the main body of the jet may be less so.

\setcounter{figure}{7}
\begin{figure}
\centering \epsfig{file=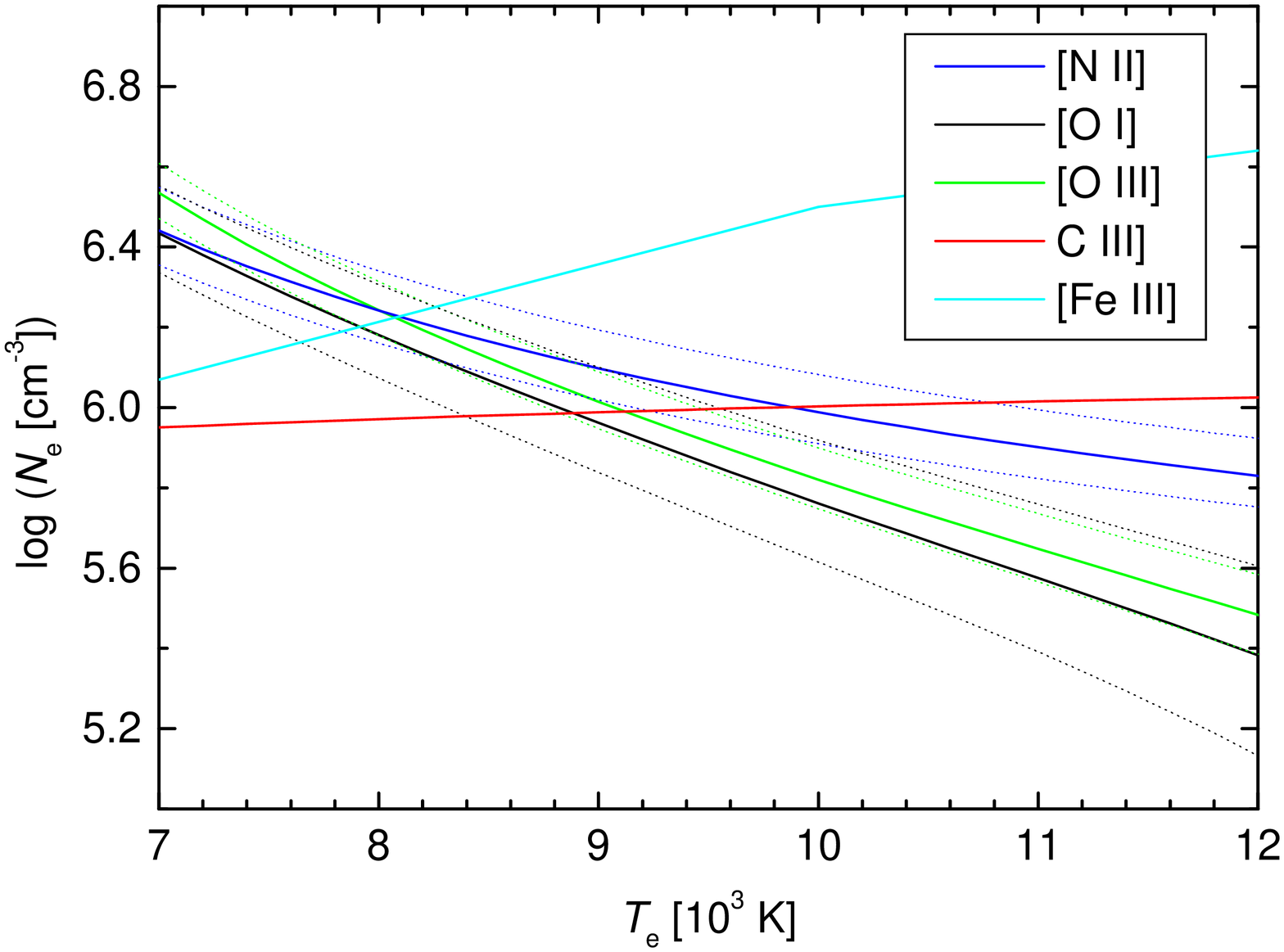, width=7 cm, scale=, clip=, angle=-90}
\caption{Electron temperature and density solutions for the red-shifted lobe of
LV\,2's jet. Dotted lines bracketing thick solid lines of the same colour
correspond to 1$\sigma$ min/max values of the respective diagnostic ratio. The
\fciii\ curve is from the {\it HST} STIS data of Henney et al. (2002) discussed
in the text.}
\end{figure}

\setcounter{figure}{8}
\begin{figure}
\centering \epsfig{file=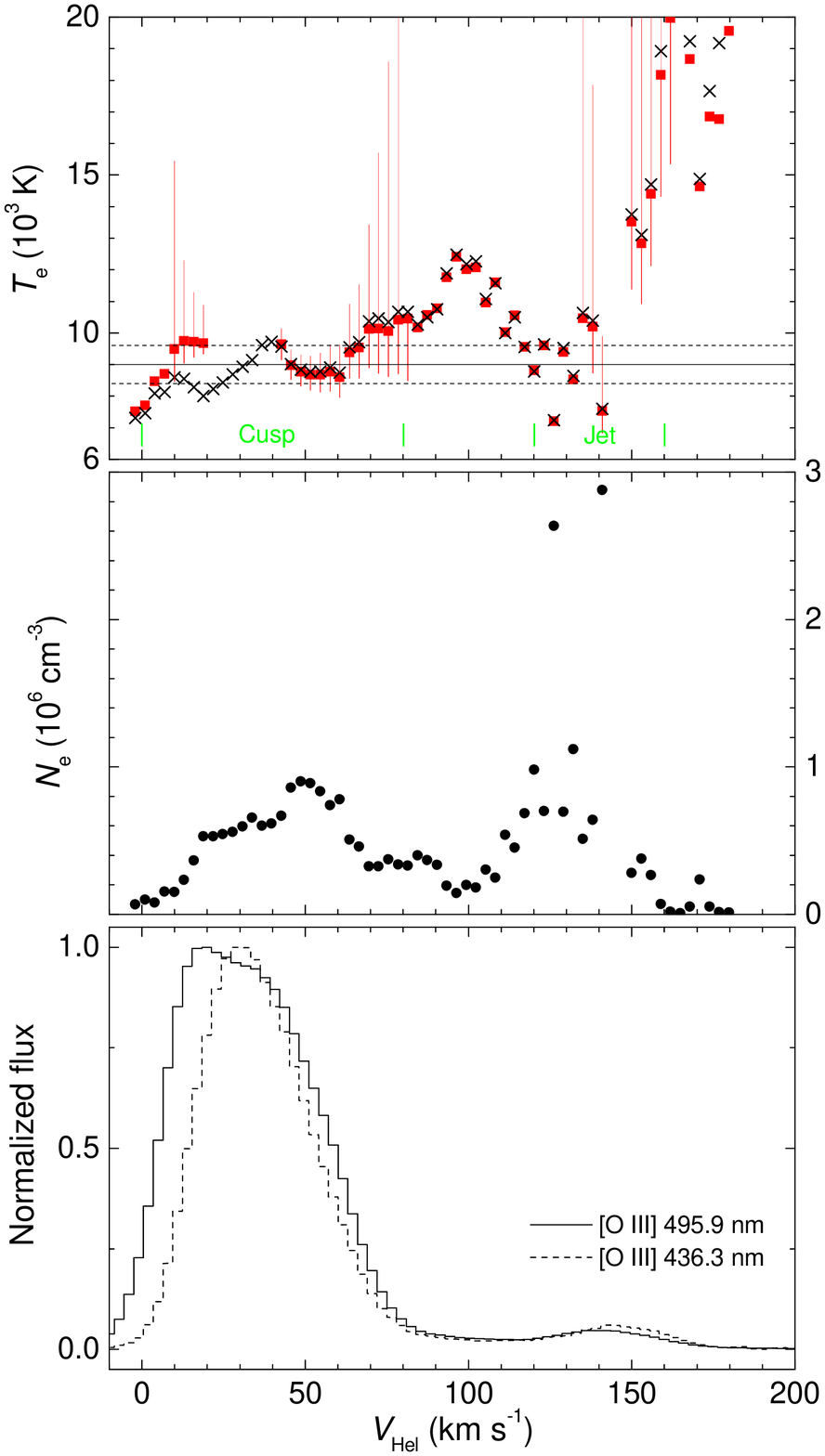, width=9.5 cm, scale=, clip=, angle=0}
\caption{The electron temperature (top) and electron density (middle)
distribution in velocity space along the line of sight towards LV\,2, based on
background-subtracted \foiii\ and \fciii\ observations (as shown in the bottom
panel and discussed in the text respectively). The solid/dashed horizontal
lines in the top panel denote the temperature and 1$\sigma$ uncertainty
measured from the integrated \foiii\ line ratio for the cusp in Paper~I (table
3).}
\end{figure}

Using the information obtained above for the physical conditions in the
red-shifted jet lobe we have computed ionic abundances for gas temperatures of
9000 and 10\,000\,K, adopting an electron density of 10$^6$ \cmt. The results
are presented in Table~4. Both dominant ionization states of sulphur are
observed allowing us to obtain its gas-phase abundance. We have assumed an
ionization correction factor ({\it icf}) of 1.05 meaning that higher ionization
stages probably contribute less than five per cent of the total (eq. 5, Paper
I). This {\it icf} is applicable to the cusp of LV\,2 and can be considered to
be a valid assumption for the jet also given the very similar ion temperature
and density in the two. The abundance of sulphur is then 12$+$log(S/H) $=$ 6.93
$\pm$ 0.06, that is, $\approx$0.2 dex less than in the Sun. This is in good
agreement with the abundance of 6.83 measured from the rest velocity component
of the lines due to emission from the cusp (table 7, Paper~I). An independent
determination for the gas-phase S/H ratio in M42 is 6.89$\pm$0.02 based on {\it
SST} data (Rubin et al. 2011).

\subsubsection{The Fe abundance and dust considerations}

Of particular interest is the abundance of \fepp/\hp\ in the jet which places a
lower limit to the total iron abundance. We used a 25-level ionic model for
\feiii\ employing the transition probabilities of Nahar \& Pradhan (1996) and
the collision strengths of Zhang (1996) to compute the abundance of \fepp\
using the $\lambda\lambda$4658, 5011 lines, finding 12 $+$ log(\fepp/\hp) $=$
7.01$\pm$0.05. This is 2 dex higher than the \emph{total} Fe/H abundance in the
cusp of LV\,2 (Paper I). It is therefore established that (i) iron in the main
body of LV\,2 is depleted by a factor of \emph{at least} 100 with respect to
that in the jet, and (ii) there is a factor of \emph{at least} nine more iron
in the jet ($\gtrsim$1 dex) than in the bulk M42 gas (cf. Paper I and Esteban
et al. 2004). The jet therefore has an iron abundance of $>$30 per cent solar
or a depletion relative to solar [Fe/H] $=$ log(Fe/H) $-$
log(Fe/H)$_{\bigodot}$ $>$ $-$0.5 dex. If we adopt the same icf(Fe) for the jet
as that for the core from Paper I then [Fe/H]$_{\rm jet}$ $\sim$ $-$0.2 dex.

An increase in the gaseous Fe abundance associated with the high-velocity
HH~202 outflow in the Orion nebula has also been attributed to dust destruction
by Mesa-Delgado et al. (2009), who found $-$(0.3--0.5) dex of iron depletion
with respect to solar. The depletion of iron in the core of LV\,2 sampled by
the rest velocity line component is [Fe/H]$_{\rm core}$ $=$ $-$(2.54$\pm$0.40)
dex (Paper~I). This is higher than the depletions of (i) $-$1.3 dex associated
with the low-velocity component of the HH~202 flow (Mesa-Delgado et al. 2009),
and (ii) $-$1.5 dex measured in the local background of LV\,2 which samples the
generic M42 gas (Paper~I). The iron depletion in LV\,2's rest frame
(0.003--0.007 of the solar abundance) is also higher than in typical Galactic
\hii\ regions, and lies within the high end of the range exhibited by some
planetary nebulae (Delgado Inglada et al. 2009). In terms of the general
interstellar medium this level of depletion is comparable only to what is found
towards a minority of sight lines, including the direction to the star
$\zeta$~Oph which represents the prototype of strong depletions observed in the
cold ISM (Jenkins 2009).

The fact that sulphur, which is a volatile element [its 50 per cent
condensation temperature, $T_{\rm C}$, into troilite (FeS) is 664\,K; Lodders
2003], shows essentially the same abundance in the core and jet of LV\,2,
whereas the refractory species iron ($T_{\rm C}$ $=$ 1334\,K) is much more
abundant in the jet than in the core leads to the following conclusions: (i)
iron-bearing condensates must be very efficiently destroyed in the fast jet,
probably as a result of grain sputtering processes, thus substantially
enhancing the overall gaseous abundance of Fe (factor of $\times$220 increase
over LV\,2's rest frame, and $\times$19 over M42); (ii) there is no evidence
for a significant depletion of sulphur onto dust and therefore LV\,2 has a
genuinely lower S abundance than the Sun ($-$0.2 dex).


\subsubsection{The Ne/S abundance ratio}

The Ne/S abundance ratio in LV\,2 is well constrained via the proxy ionic ratio
\nepp/\sulpp\ as it involves lines of similar excitation energy (37.2\,kK for
\fneiii\ $\lambda$3869 and 39.1\,kK for \fsiii\ $\lambda$6312) and critical
density (7.1$\times$10$^6$ vs. 1.0$\times$10$^7$ \cmt), and is therefore fairly
insensitive to the assumption of \elt\ or \eld\ for their emitting zones. This
ratio is 13.5 and 14.8 in the jet and core of LV\,2 respectively (cf. Table 4
with Paper I, table 5) whereas the canonical solar Ne/S ratio is 6.5 (Asplund
et al. 2009). Our measurement for LV\,2 agrees with the value of 13.0
applicable to M42 from infrared lines measured by \textit{Spitzer} (Rubin et
al. 2011). The IR lines set a good benchmark as their are immune to \elt\ or
reddening biases. It is discomfiting however that the ratio in the local
vicinity of LV\,2 measured using the optical lines is only 4.9 (Paper I, table
5), and in agreement with the value of 4.8 from Esteban et al. (2004)
applicable to a different M42 zone obtained also from the optical lines.

Two factors could conspire to bring about a discrepancy between the optical and
IR values of the Ne/S ratio for M42: in terms of \elt\ effects, the offset
could be explained if the optical \nepp/\sulpp\ ratio was underestimated due to
a variation of $\sim$1900\,K between a cooler \nepp\ and a warmer \sulpp\
emitting volume (rather than assuming the same temperature for both). Secondly,
the differential reddening correction between 3869 and 6312\,\AA\ is a factor
of about two, and if the extinction along the optical observation sightlines
has been underestimated in any way this would also cause the optical
\nepp/\sulpp\ ratio to appear too low. On the other hand, it would be very hard
to appeal to \elt\ (or extinction) biases to explain the opposite, i.e. that
the Ne/S ratio in LV\,2 has been overestimated, as this would have to act
opposite to the natural tendency of the $\lambda$3869 and $\lambda$6312 lines
to be more efficiently excited at lower and higher \elt, respectively.

We therefore conclude that the Ne/S ratio in LV\,2 is genuinely higher than
solar by factors of 2.1 -- 4.3. The lower value is obtained from the
\nepp/\sulpp\ determination applicable to the jet and core of LV\,2. The upper
limit is obtained from the total Ne/S ratio for the rest velocity component
(core) from Paper I and may be affected by ionization correction factor
considerations. This adds to the body of evidence that the canonical solar Ne
(Asplund et al. 2009) appears to be low compared to M42 (Rubin et al. 2011) and
the LV\,2 proplyd (Paper I and this work) based on considerations of the Ne/S
ratio, the absolute neon abundance in B-type stars in Orion (S\'{i}mon-D\'{i}az
\& Stasi\'{n}ska 2011 and references therein), or when comparing the solar Ne/O
ratio to the mean value from Galactic planetary nebulae and \hii\ regions (e.g.
Wang \& Liu 2008). A new solar abundance model by Delahaye et al. (2011)
partially rectifies this situation by raising the Ne abundance to 8.15$\pm$0.17
($+$0.22 dex with respect to the value in Asplund et al 2009).

\begin{table}
\caption{Fluxes of emission lines from the red-shifted jet of LV\,2 in units
such that \hb\ $=$ 100.$^a$}
\begin{tabular}{lcc}
\noalign{\vskip3pt} \noalign{\hrule} \noalign{\vskip3pt}
Line              &$F$($\lambda$)  &$I$($\lambda$)              \\
\noalign{\vskip3pt} \noalign{\hrule} \noalign{\vskip3pt}
\fneiii\ $\lambda$3869     &45.1 $\pm$ 6.45                          &67.2 $\pm$ 9.7             \\
\fsii\ $\lambda$4069       &5.15 $\pm$ 0.71                          &7.19 $\pm$ 0.99             \\
\hd\ $\lambda$4101         &20.4$\pm$ 0.9                            &27.9$\pm$  1.6                 \\
\hg\ $\lambda$4340         &32.8 $\pm$ 2.6                           &42.3$\pm$  3.4                 \\
\foiii\ $\lambda$4363      &5.93 $\pm$ 0.55                          &7.44 $\pm$ 0.70             \\
\hei\ $\lambda$4471        &5.21 $\pm$ 0.55                          &6.22 $\pm$ 0.66             \\
\ffeiii\ $\lambda$4658     &12.4 $\pm$ 0.94                          &13.0 $\pm$ 1.0             \\
\ffeiii\ $\lambda$4702     &5.56 $\pm$ 0.57                          &5.9 $\pm$  0.61             \\
\ffeiii\ $\lambda$4734     &3.03 $\pm$ 0.39                          &3.20 $\pm$ 0.41             \\
\hb\ $\lambda$4861         &100 $\pm$ 4                              & 100$\pm$  5               \\
\foiii\ $\lambda$4959      &103  $\pm$ 9.4                           &99.0$\pm$  9.0             \\
\foiii\ $\lambda$5007      &309 $\pm$ 29                             &290$\pm$   28            \\
\ffeiii\ $\lambda$5011     &4.75 $\pm$ 0.72                          &4.46$\pm$  0.68             \\
\foi\ $\lambda$5577        &0.231$\pm$ 0.046                         &0.174$\pm$ 0.035            \\
\fnii\ $\lambda$5755       &6.97 $\pm$ 1.16                          &4.95$\pm$  0.83             \\
\foi\ $\lambda$6300        &9.10 $\pm$ 0.49                          &5.39$\pm$  0.28             \\
\fsiii\ $\lambda$6312      &6.99 $\pm$ 0.50                          &4.14$\pm$  0.30             \\
\ha\ $\lambda$6563         &287 $\pm$ 20:                            &201$\pm$   14:       \\
\fnii\ $\lambda$6584       &7.21 $\pm$ 0.43                          &3.90$\pm$  0.25             \\

\noalign{\vskip3pt} \noalign{\hrule}\noalign{\vskip3pt}
\end{tabular}
\begin{description}
\item[$^a$] Several lines were measured on the lower resolution Argus spectra
using the gratings quoted in table 1 of Paper~I: \fneiii\ $\lambda$3869 (LR1),
\hei\ $\lambda$4471 (LR2), \hb\ $\lambda$4861 (LR3), \foi\ $\lambda$5577 (LR4),
\fnii\ $\lambda$5755 (LR4/LR5), \foi\ $\lambda$6300 (LR5), \fsiii\
$\lambda$6312 (LR5). The values for \ha\ are unreliable. See the text for
details.
\end{description}
\end{table}

\begin{table}
\caption{Abundances for the red-shifted jet lobe of LV\,2 (in a scale where
log\,H $=$ 12 for the first seven rows and in decimal units for the last three
rows).$^a$}
\begin{tabular}{lcc}
\noalign{\vskip3pt} \noalign{\hrule} \noalign{\vskip3pt}
Species                 &\elt\ $=$ 9000\,K    &\elt\ $=$ 10000\,K              \\
                    &\eld\ $=$ 10$^6$\,\cmt &\eld\ $=$ 10$^6$\,\cmt     \\
\noalign{\vskip3pt} \noalign{\hrule} \noalign{\vskip3pt}

\np/\hp\                &7.70$^{+0.07}_{-0.08}$               &7.46$^{+0.07}_{-0.08}$ \\ 
\opp/\hp\               &8.60$^{+0.04}_{-0.05}$               &8.29$^{+0.03}_{-0.05}$ \\ 
\nepp/\hp\              &8.05$^{+0.10}_{-0.03}$               &7.85$^{+0.10}_{-0.03}$ \\
\fepp/\hp\              &7.08$^{+0.03}_{-0.03}$               &6.92$^{+0.04}_{-0.03}$    \\ 
\sulp/\hp\              &6.24$^{+0.05}_{-0.07}$               &6.07$^{+0.08}_{-0.06}$ \\
\sulpp/\hp\             &6.92$^{+0.04}_{-0.04}$               &6.71$^{+0.05}_{-0.03}$ \\
S/H                     &7.02$^{+0.04}_{-0.04}$               &6.82$^{+0.05}_{-0.05}$ \\ 
\noalign{\vskip3pt}
\fepp/\opp\         &0.030$\pm$0.004              &0.043$\pm$0.006  \\
\nepp/\opp\         &0.28$\pm$0.08               &0.36$\pm$0.10 \\
\nepp/\sulpp\       &13.5$\pm$3.7               &13.8$\pm$3.8   \\

\noalign{\vskip3pt} \noalign{\hrule}\noalign{\vskip3pt}
\end{tabular}
\begin{description}
\item[$^a$] Solar abundances for comparison: N $=$ 7.83$\pm$0.05, O $=$
8.69$\pm$0.05, Ne $=$ 7.93$\pm$0.10, S $=$ 7.12$\pm$0.03, and Fe $=$
7.50$\pm$0.04 (Asplund et al. 2009).
\end{description}
\end{table}

\section{Discussion and conclusions}

We have presented spatially resolved, high dispersion spectra of LV\,2 and its
Orion nebula vicinity based on VLT observations. The deep spectra enabled the
detection and resolution of emission lines arising from the proplyd's
irradiated bipolar microjet. Maps of LV\,2 in the light of \hi\ and \ffeiii\
lines have been presented, where the jet lobes appear spatially resolved from
the core of the proplyd.

An abundance analysis was undertaken for the brighter red-shifted jet lobe
complementing the abundance analysis of the core of LV\,2 presented in Paper~I.
A noteworthy result is the significantly enhanced abundance of iron in the jet
versus the core of LV\,2, pointing to very efficient dust destruction
mechanisms in fast irradiated microjets. The overall depletion of iron in the
proplyd's rest frame is higher than in Galactic \hii\ regions and is comparable
to what is found along cold ISM sight lines. The Ne/S abundance ratio in LV\,2
as probed by optical lines is $\geq$2 times solar, in agreement with the value
applicable to the generic M42 gas from IR line measurements by {\it Spitzer}.

The extreme-UV irradiated jet shows a range of ionization conditions allowing
the survival of neutral oxygen pockets mixed in with highly ionized gas. A
velocity-resolved analysis was performed probing the electron temperature and
density across LV\,2's rest frame and red-shifted jet lobe. The \foiii\ and
\fciii\ emitting regions in LV\,2 show a range of densities with a peak of
10$^6$ \cmt\ near the line systemic velocity and an upper limit of $\sim$10$^7$
\cmt\ from \ffeiii\ emission within the red-shifted jet lobe. The electron
temperature in the highly ionized zone in the main body of the proplyd shows a
small 10 per cent rms variation. Larger temperature variations associated with
the onset of the jet emission have been attributed to a shock discontinuity.

The far-UV to optical collisional lines emitted from LV\,2 are correlatively
quenched according to the critical densities of their upper levels, with lines
of low critical density being affected the most. This reinforces the conclusion
of Paper~I that in \hii\ regions containing dense zones, such as proplyds,
filaments, globules etc., line ratios such as \foiii\
$\lambda$4363/($\lambda$4959$+$$\lambda$5007) and \fnii\
$\lambda$5755/($\lambda$6548$+$$\lambda$6584) will be heavily weighted towards
the dense component of the gas. These line ratios are the main temperature
diagnostics for Galactic and extragalactic \hii\ regions, whose validity rests
on an independent way of determining the mean density of the plasma. This is
usually done via the \foii\ or \fsii\ doublet ratios which are demonstrably
biased towards low density regions (Fig.\,~5 left). The \fnii\ ratio will be
especially biased as the critical densities of its constituent lines are at a
ratio of approximately 200 versus only 40 for the \foiii\ lines, resulting in
more uneven fractions of \fnii\ $\lambda$5755 versus \fnii\ $\lambda$6584 to be
emitted from dense and diffuse gas respectively, than is the case for \foiii\
$\lambda$4363 and \foiii\ $\lambda$5007 -- this is demonstrated in Fig.\,~5.

In this situation, and especially in cases of distant nebulae whose gas density
distribution appears deceptively homogeneous on large scales, or in the
presence of condensations just below the angular resolution limit in nearby
nebulae, the end result would be an overestimation of the electron temperature
and an inevitable underestimation of the gas metallicity (e.g. Rubin 1989).
Given the behaviour of the \fnii\ and \foiii\ lines seen in Fig.\,~5 the former
temperature would be more adversely affected that the latter in \hii\ regions
with a clumpy small-scale density distribution. It may be no coincidence
therefore that \elt(\fnii) values in \hii\ regions are typically higher than
\elt(\foiii) values by $\simeq$1000\,K on average (e.g. Rodr\'iguez \&
Delgado-Inglada 2011). This is usually justified in the literature as being due
to the fact that the low ionization zones in a nebula tend to have a higher
temperature than the \oiii\ zone. Based, however, on the arguments presented
here this behaviour could be at least partly spurious if the underlying density
distribution of a nebula is sufficiently clumpy at sub-arcsecond scales.

Regarding the abundance anomaly pertaining to RL versus CL diagnostics (e.g.
Tsamis et al. 2003; Tsamis \& P\'equignot 2005; Garc\'ia-Rojas \& Esteban 2007;
Ercolano 2009; Esteban et al. 2009; Mesa-Delgado et al. 2010;
Sim{\'o}n-D{\'{\i}}az \& Stasi{\'n}ska 2011; Paper~I), it should be noted that
the fraction of \oii\ $\lambda$4649, \cii\ $\lambda$4267 and \hi\ RLs that are
emitted from LV\,2 is the same for all these lines ($\approx$70 per cent),
while the rest is emitted from diffuse M42 gas along the line of sight. As a
result the proplyd does not show elevated \opp/\hp\ and \cpp/\hp\ abundance
ratios, measured using RLs, compared to the Orion background (cf. Paper~I). In
our view this analysis goes some way towards explaining the abundance anomaly
in \hii\ regions as a problem affected by the use and interpretation of the
\foiii\ and \fnii\ plasma thermometers when applied to average spectra of large
nebular volumes.

\section*{Acknowledgments}

We thank the FLAMES support astronomers at ESO for scheduling the VLT service
mode observations. William Henney kindly gave us access to reduced {\it HST}
STIS data of LV\,2. We thank Bob Rubin, Daniel P\'equignot and Adal
Mesa-Delgado for useful correspondence. The referee is thanked for a detailed
and very helpful report. This research made use of the NASA ADS database.

YGT acknowledges the award of a Marie Curie intra-European Fellowship within
the 7$^{\rm th}$ European Community Framework Programme (grant agreement
PIEF-GA-2009-236486).

\label{lastpage}


\begin{thebibliography}{}

\bibitem[Adams(2010)]{} Adams, F. C.\ 2010, ARA\&A, 48, 47

\bibitem[Asplund et al.(2009)]{} Asplund, M., Grevesse, N., Jacques Sauval, A., \& Scott, P.\ 2009, ARA\&A, 47,
481

\bibitem[Barlow(2006)]{} Barlow M. J., Hales A. S., Storey P. J., Liu X.-W., Tsamis Y. G., Aderin
M. E., 2006, in Barlow M. J., M´endez R. H., eds, Proc. IAU Symp. 234,
Planetary Nebulae in our Galaxy and Beyond. Cambridge Univ. Press, Cambridge,
p. 367

\bibitem[Boss(2010)]{2010IAUS..265..391B} Boss, A.~P.\ 2010, IAU Symposium,
265, 391

\bibitem[Blagrave et al.(2007)]{2007ApJ...655..299B} Blagrave, K.~P.~M.,
Martin, P.~G., Rubin, R.~H., Dufour, R.~J., Baldwin, J.~A., Hester, J.~J., \&
Walter, D.~K.\ 2007, \apj, 655, 299

\bibitem[Blecha \& Simond 2004]{} Blecha, A., \& Simond, G.\ 2004, GIRRAFE BLDR
Software Reference Manual 1.12 (http://girbldrs.sourceforge.net)

\bibitem[Cardelli et al.(1989)]{1989ApJ...345..245C} Cardelli, J.~A.,
Clayton, G.~C., \& Mathis, J.~S.\ 1989, \apj, 345, 245 (CCM)

\bibitem[Davey et al.(2000)]{2000A&AS..142...85D} Davey, A.~R., Storey,
P.~J., \& Kisielius, R.\ 2000, \aap Supplement, 142, 85

\bibitem[Delahaye et al.(2010)]{2010arXiv1005.0423D} Delahaye, F.,
Pinsonneault, M.~H., Pinsonneault, L.,
\& Zeippen, C.~J.\ 2010, arXiv:1005.0423

\bibitem[Delgado Inglada et al.(2009)]{2009ApJ...694.1335D} Delgado
Inglada, G., Rodr{\'{\i}}guez, M., Mampaso, A.,
\& Viironen, K.\ 2009, \apj, 694, 1335

\bibitem[Ercolano(2009)]{2009MNRAS.397L..69E} Ercolano, B.\ 2009, \mnras,
397, L69

\bibitem[Esteban \& Peimbert(1999)]{} Esteban, C., Peimbert,
M.\ 1999, A\&A, 349, 276

\bibitem[Esteban et al.(2004)]{2004MNRAS.355..229E} Esteban, C., Peimbert,
M., Garc{\'{\i}}a-Rojas, J., Ruiz, M.~T., Peimbert, A., \& Rodr{\'{\i}}guez,
M.\ 2004, \mnras, 355, 229

\bibitem[Esteban et al.(2009)]{2009ApJ...700..654E} Esteban, C., Bresolin,
F., Peimbert, M., Garc{\'{\i}}a-Rojas, J., Peimbert, A., \& Mesa-Delgado, A.\
2009, \apj, 700, 654

\bibitem[Garc\'ia-Rojas \& Esteban 2007]{} Garc\'ia-Rojas, J. \& Esteban, C.\
2007, \apj, 670, 457

\bibitem[Garc{\'{\i}}a-D{\'{\i}}az et al.(2008)]{2008RMxAA..44..181G}
Garc{\'{\i}}a-D{\'{\i}}az, M.~T., Henney, W.~J., L{\'o}pez, J.~A., \& Doi, T.\
2008, RevMexAA, 44, 181

\bibitem[Gonzalez(1997)]{1997MNRAS.285..403G} Gonzalez, G.\ 1997, \mnras,
285, 403

\bibitem[Hamuy(1992)] {} Hamuy, M., Walker, A.R., Suntzeff, N.B., et al., 1992,
PASP, 104, 533

\bibitem[Hamuy(1994)] {} Hamuy, M., Suntzeff, N.B., Heathcote, S.R.,  et al., 1994,
PASP, 106, 566

\bibitem[Henney \& O'Dell(1999)]{1999AJ....118.2350H} Henney, W.~J., \& O'Dell, C.~R.\ 1999, \aj, 118, 2350

\bibitem[Henney et al.(2002)]{2002ApJ...566..315H} Henney, W.~J., O'Dell,
C.~R., Meaburn, J., Garrington, S.~T., \& Lopez, J.~A.\ 2002, \apj, 566, 315

\bibitem[Jenkins(2009)]{2009ApJ...700.1299J} Jenkins, E.~B.\ 2009, \apj,
700, 1299

\bibitem[Johnstone et al.(1998)]{1998ApJ...499..758J} Johnstone, D.,
Hollenbach, D., \& Bally, J.\ 1998, \apj, 499, 758

\bibitem[Keenan et al.(2001)]{2001PNAS...98.9476K} Keenan, F.~P.,
Aller, L.~H., Ryans, R.~S.~I., \& Hyung, S.\ 2001, Proceedings of the National
Academy of Science, 98, 9476

\bibitem[Laques \& Vidal(1979)]{1979A&A....73...97L} Laques, P., \& Vidal, J.~L.\ 1979, \aap, 73, 97

\bibitem[Lodders(2003)]{2003ApJ...591.1220L} Lodders, K.\ 2003, \apj, 591,
1220

\bibitem[Mesa-Delgado et al.(2009)]{2009MNRAS.395..855M} Mesa-Delgado, A.,
Esteban, C., Garc{\'{\i}}a-Rojas, J., Luridiana, V., Bautista, M.,
Rodr{\'{\i}}guez, M., L{\'o}pez-Mart{\'{\i}}n, L., \& Peimbert, M.\ 2009,
\mnras, 395, 855

\bibitem[Mesa-Delgado \& Esteban(2010)]{} Mesa-Delgado, A., \& Esteban, C.\ 2010,
MNRAS, 405, 2651

\bibitem[Nahar \& Pradhan(1996)]{} Nahar \& Pradhan \ 1996, A\&AS, 119, 509

\bibitem[Neves et al.(2009)]{2009A&A...497..563N} Neves, V., Santos, N.~C., Sousa, S.~G., Correia, A.~C.~M., \& Israelian, G.\ 2009, \aap, 497, 563


\bibitem[O'Dell et al.(1993)]{1993ApJ...410..696O} O'Dell, C.~R., Wen, Z., \& Hu, X.\ 1993, \apj, 410, 696

\bibitem[O'dell(2001)]{2001ARA&A..39...99O} O'Dell, C.~R.\ 2001, ARA\&A, 39, 99

\bibitem[Oke 1990]{} Oke, J.B.\ 1990, AJ, 99, 1621

\bibitem[Reid et al.(2009)]{2009ApJ...700..137R} Reid, M.~J., et al.\ 2009, \apj, 700, 137

\bibitem[Rodr\'iguez \& Delgado-Inglada(2011)]{} Rodr\'iguez, M. \& Delgado-Inglada, G.\ 2011, ApJL, 733,
L50

\bibitem[Rubin(1989)]{1989ApJS...69..897R} Rubin, R.~H.\ 1989, \apjs, 69, 897

\bibitem[Rubin et al.(2011)]{} Rubin, R.~H., Simpson,
J.~P., O'Dell, C.~R., McNabb, I.~A., Colgan, S.~W.~J., Zhuge, S.~Y., Ferland,
G.~J., \& Hidalgo, S.~A.\ 2011, MNRAS, 410, 1320


\bibitem[Sim{\'o}n-D{\'{\i}}az(2010)]{2010A&A...510A..22S} Sim{\'o}n-D{\'{\i}}az, S.\ 2010, \aap, 510, A22

\bibitem[Sim{\'o}n-D{\'{\i}}az \& Stasi{\'n}ska(2011)]{2011A&A...526A..48S} Sim{\'o}n-D{\'{\i}}az, S., \& Stasi{\'n}ska, G.\ 2011, \aap, 526, A48

\bibitem[Storey \& Hummer(1995)]{1995MNRAS.272...41S} Storey, P.~J., \& Hummer, D.~G.\ 1995, \mnras, 272, 41

\bibitem[Tsamis et al.(2003)]{2003MNRAS.338..687T} Tsamis, Y.~G., Barlow, M.~J., Liu, X.-W., Danziger, I.~J., \& Storey, P.~J.\ 2003a, \mnras, 338, 687

\bibitem[Tsamis \& P{\'e}quignot(2005)]{2005MNRAS.364..687T} Tsamis, Y.~G., \& P{\'e}quignot,
D.\ 2005, \mnras, 364, 687

\bibitem[Tsamis et al.(2011)]{2011MNRAS.412.1367T} Tsamis, Y.~G., Walsh,
J.~R., V{\'{\i}}lchez, J.~M., \& P{\'e}quignot, D.\ 2011, \mnras, 412, 1367
(Paper~I)

\bibitem[Vasconcelos et al.(2005)]{2005AJ....130.1707V} Vasconcelos, M.~J., Cerqueira, A.~H., Plana, H., Raga, A.~C., \& Morisset, C.\ 2005, \aj, 130, 1707

\bibitem[Wang \& Liu(2008)]{2008MNRAS.389L..33W} Wang, W., \& Liu, X.-W.\ 2008, \mnras, 389, L33

\bibitem[Zhang(1996)]{} Zhang\ 1996, A\&AS, 119, 523

\end{thebibliography}
\end{document}